\documentclass[12pt,preprint]{aastex}

\def\msun{$M_{\odot}$}
\def\rsun{$R_{\odot}$}
\def\kms{km s$^{\rm -1}$}

\begin{document}

\title{Line Broadening in Field Metal-poor
Red Giant and Red Horizontal Branch Stars}

\author{Bruce W. Carney}
\affil{Department of Physics \& Astronomy\\University of North Carolina\\
Chapel Hill, NC 27599-3255\\e-mail: bruce@unc.edu\\}
\author{David W. Latham \& Robert P. Stefanik}
\affil{Harvard-Smithsonian Center for Astrophysics\\60 Garden Street,
Cambridge, MA  02138\\e-mail: dlatham@cfa.harvard.edu; rstefanik@cfa.harvard.edu\\}
\author{John B. Laird}
\affil{Department of Physics \& Astronomy\\Bowling Green State University\\
Bowling Green, OH 43403\\e-mail: laird@bgsu.edu\\}

\begin{abstract}
We report
349 radial velocities for 45 metal-poor field
red giant and red horizontal branch stars, with time coverage ranging
from 1 to 21 years. We have identified one new spectroscopic
binary, HD~4306, and one possible such system, HD~184711.
We also report 57 radial velocities for 11 of the 91 stars reported
on previously by Carney et al.\ (2003). All but one of the 11 stars
had been found to have variable radial velocities. New velocities
for the long-period spectroscopic binaries BD$-1$~2582 and
HD~108317 have extended the time coverage to 21.7 and 12.5 years,
respectively, but in neither case have we yet completed a full
orbital period.
As was found in the previous study, radial
velocity ``jitter" is present in many of the most luminous stars.
Excluding stars showing spectroscopic binary orbital motion, all 7
of the red giants
with estimated $M_{\rm V}$ values more luminous than $-2.0$ display jitter,
as well as 3 of the 14 stars with $-2.0 < M_{\rm V} \leq\ -1.4$. 
We have also measured the line broadening in all the new spectra,
using synthetic spectra as templates. Comparison with results from high-resolution
and higher signal-to-noise (S/N) spectra 
employed by other workers shows good agreement
down to line broadening levels of 3 \kms, well below our instrumental
resolution of 8.5 \kms. As the previous work demonstrated, most of the
most luminous red giants show significant line broadening, as do many
of the red horizontal branch stars, and we discuss briefly possible
causes. The line broadening appears related to velocity jitter,
in that both appear primarily among the highest luminosity
red giants.

\end{abstract}


\section{INTRODUCTION}

In a previous paper (Carney et al.\ 2003 hereafter C2003), we
discussed results from over two thousand high-resolution
(R = 32,000), low signal-to-noise (10 to 50 per resolution
element) spectra of 91 field metal-poor red giant branch (RGB) and red horizontal
branch (RHB) stars. Radial velocities were obtained by cross-correlating
each observed spectrum with a synthetic spectrum that closely
matched the adopted temperature, gravity, and metallicity for
the star.
Sixteen stars were found to be single-lined
spectroscopic binaries, and orbital solutions were presented
for 14 of them. Excluding those 14 stars, observations
of the stars covered spans from 2956 days to 6670 days,
roughly from 8 to 18 years, with an average of 13.7 years.

The use of synthetic spectra
enabled us to measure line broadening as well as 
radial velocities. We found some
anticipated results as well as some surprises. For example,
studies of RGB stars in globular clusters (Gunn \& Griffin 1979; 
Mayor \& Mermilliod 1984; Lupton et al.\ 1987;
Pryor et al.\ 1988; C\^{o}t\'{e} et al.\ 1996, 2002;
Brown \& Wallerstein 1992; 
Kraft et al.\ 1997; Mayor et al.\ 1997)
have shown that the most luminous stars, generally with
$M_{\rm V} \leq\ -1.4$, show velocity variability that is
unlikely to be associated with orbital motion. About half
of the stars studied by C2003 with estimated $M_{\rm V}$ values
more luminous than $-1.4$ also showed such velocity variability, which is
often referred to as ``jitter". The term's ambiguity
describes the absence of a well-understood
cause of the velocity variability.

As expected,
the spectroscopic binary frequency, for periods less than 6000 days,
was very similar to that found for metal-poor dwarfs and subgiants
(Latham et al.\ 2002; Goldberg et al.\ 2002). There was some
evidence for a dearth of short-period binaries among the
RGB and RHB stars, which is not too surprising.
Most short period systems are expected to undergo
mass transfer when the more massive star begins to expand and
fills its Roche lobe. The period does shorten initially,
until the donor star's mass becomes less than that of the
original secondary star, following which the orbital
separation widens and the period increases.
Among the fourteen spectroscopic
binaries in C2003 with orbital solutions, two systems 
with small ratios of orbital semi-major axis to estimated
primary stellar radius have circular orbits, presumably the result of 
tidal interactions. These two stars also had higher line broadenings than
other RGB stars,
presumably due to higher rotational velocities, 
a consequence of tidally-induced locking
of the rotational and orbital periods\footnote{C2003 referred to
line broadening as rotational velocities since the values were
determined using rotationally-broadened synthetic spectra. Macroturbulence
also contributes to line broadening, so throughout this paper we
refer to the more general term, line broadening, and refer
to $V_{\rm broad}$ rather than $V_{\rm rot}$~sin~$i$.}.

The most unexpected result involved the dependence of the
measured line broadening on the stars' evolutionary state. 
For most of the RGB stars, the line broadening was generally
smaller than our instrumental resolution (about 8.5 \kms).
But at the highest luminosities, the mean line broadening
rose to values as high as about 12 \kms\ at $M_{\rm V} \approx\ -2.6$,
and perhaps as high as 15 \kms\ if our interpretation of the
apparent periodicity in the velocity jitter of two stars was
interpreted correctly as due to the rotational period. These very
high levels of line broadening seemed to rule out macroturbulence.
We also discounted transfer of core rotational angular momentum
to the surface of the RGB stars because we should have
seen significant increases in line broadening where internal
mixing manifests itself by changes in atmospheric abundances.
Having ruled out alternative explanations, we speculated that
a ``spin up" in outer envelope rotational velocities could
have been caused by the absorption of giant planets.
This would imply that metal-poor stars might have such
planets {\em and} that they exist at larger orbital distances
from the host stars than have been found to date in 
many metal-rich disk stars.

We also observed significant line broadening in the RHB stars.
To first order, this makes
sense. If the much larger stars near the tip of the
RGB stars rotate, then so should their
descendents. But perhaps their descendents should be
stars whose enhanced rotation helps them shed their
outer envelopes, which should lead to core helium-burning
stars blueward of the instability strip, known as blue horizontal
branch (BHB) stars. Soker (1998) and Siess \& Livio (1999)
argued that absorption of a planet by a luminous red
giant could explain the high rotation velocities often
found among field BHB stars (Kinman et al.\ 2000; Behr 2003)
and cluster BHB stars (Peterson et al.\ 1983; Peterson 1985a;
Peterson et al.\ 1995; Cohen \& McCarthy 1997; Behr et al.\ 2000a,b;
Recio-Blanco et al.\ 2002). More generally, theorists have
struggled to explain the presence of both RHB and BHB stars
in clusters whose stars share the same metallicity. Some ``second parameter"
must be at work, presumably leading to either larger (RHB) envelope
masses or smaller ones (BHB stars). If rotation is the
second parameter, then why do RHB stars in the field appear
to have rotational velocities comparable to BHB stars,
when allowance is made for their different radii?

C2003 suggested four follow-up studies. First,
expand the sample to ascertain if our
original sample was unusual in some manner. This paper
reports on the results of a ``hasty" study of 45
additional metal-poor field stars\footnote{By ``hasty", we mean that
some stars were observed for less than two years.}. 
We also sought to obtain line broadening measures for giants in
globular clusters, and initial data for four clusters are
in hand. Results will be published later. If planets
do exist around metal-poor stars, this would suggest
that disk instability is a viable mechanism for planet
formation, and a high-precision radial velocity
survey of roughly two hundred metal-poor field stars
was begun (see Sozzetti et al.\ 2006).
Finally, we have to ask if the line broadening is due
to rotation or to macroturbulence. This can be
determined with very high-resolution, high-S/N
spectra that enable Fourier transform studies of
line profiles, following Gray (1982; 1984) and
Gray \& Toner (1986; 1987). We have completed acquisition
and analysis of such spectra and will report on the results in a
future paper (Carney et al.\ 2007).

\section{SELECTION OF STARS FOR STUDY}

C2003 discussed in detail the criteria by which they
assembled their list of RGB and RHB field stars for study.
These followed mostly the earlier study of Carney \& Latham (1986),
which in turn relied on the 
kinematically-unbiased metal-poor ([Fe/H] $\leq$\ $-1.5$) samples
identified by Bond (1980). He identifield weak-lined stars using
objective prism spectroscopy as well as follow-up
$uvby$ photometry. 
Norris, Bessell, \& Pickles (1985) undertook a large program
of additional DDO and $RI_{\rm C}$ photometry.
Most of the stars
in the C2003 sample had been classified as metal-poor
red giants by Anthony-Twarog \& Twarog (1994;
hereafter ATT), who supplied estimated reddening
and $M_{\rm V}$ values. All but one of the 45 stars discussed in this
paper come from ATT, and most of them
were identified originally by Bond (1980). The stars in this study
represent most of the stars in ATT with [Fe/H] $\leq\ -1.5$
not studied by C2003 but within reach of the telescopes employed
in that program.
To make certain
that the stars are all luminous, we have employed
the Hipparcos database, which includes 44 of the stars
(only BD+25~3410 lacks a measured trigonometric parallax).
The parallaxes are all very small and consistent with
high luminosities.

\section{OBSERVATIONS}

The spectra of our new program stars have been 
obtained in the same fashion as those studied
by C2003, using the Center for Astrophysics Digital Speedometers
(Latham 1985, 1992), primarily with
the 1.5-m Wyeth reflector at the Oak Ridge Observatory in
Harvard, Massachusetts\footnote{Alas, the Oak Ridge Observatory
is now closed, inhibiting such long-term radial velocity
work.}, as well as
the 1.5-m Tillinghast reflector and the MMT instruments
atop Mt. Hopkins in Arizona. The Tillinghast reflector was
especially important for the stars south
of $-20$\arcdeg\ and north of $+62$\arcdeg\ in declination.
Also as before, the wavelength coverage is 45~\AA, centered
near 5187 \AA, with a resolution of
8.5~\kms. The signal-to-noise ratio varied from 10 to 50
per resolution element, with a typical value of about 15.

C2003 described in detail the measurement of the radial
velocities. A grid of model atmospheres, defined by $T_{\rm eff}$,
log~$g$, and [Fe/H] values, was computed. 
The grid spacing in temperature was 250~K,
0.5 dex in log~$g$, and 0.5 dex in metallicity for [Fe/H] $\leq\ -1.0$.
functions and elemental abundances in which all the 
``$\alpha$" elements (O, Ne, Mg, Si, S, Ca, and Ti)
were enhanced by 0.4 dex relative to the solar
abundances. More details may be found in N\"{o}rdstrom et al.\
(1994) and C2003.

The program SYNTHE was used to compute synthetic spectra
in the wavelength range 5146-5229 \AA. In the case of the Sun, we
obtained an excellent line-by-line match between 
the Kurucz solar model and {\em synthetic} spectrum 
compared with the {\em observed} solar flux spectrum
(Kurucz et al.\ 1984).
SYNTHE computes
specific intensity at 17 different emergent angles
across the stellar disk, and integration over the disk,
including the effects of stellar rotation, yields the
synthetic flux spectrum. 
SYNTHE enables us to include the effects of 
instrumental resolution, which we chose to be
a Gaussian with a FWHM of 8.5~\kms, which is appropriate
to the instrumentation we employed. The adopted microturbulent
velocities were 2~\kms, and macroturbulent velocities were 3 \kms.
At each combination of temperature, gravity, and metallicity
in our grid we computed synthetic spectra employing a wide range
of rotational broadening profiles, with $V_{\rm rot}$~=~ 0, 1, 2,
4, 6, 8, 10, 12, 16, 20, 25, 30, 35, 40, 50, 60, 70, 80, 90,
100, 110, 120, and 140~\kms.

Once the stellar parameters, $T_{\rm eff}$, log~$g$, and [Fe/H] had
been estimated, we relied on the model atmosphere grid point closest
in these variables, paying special attention to the primary variable,
temperature. In the case of more than one close match, we employed
the template that gave the highest value for the peak correlation,
averaged over all the observed spectra.

\section{STELLAR PARAMETERS}

\subsection{Basic Procedure}

Estimation of stellar
atmospheric parameters follows the procedures described
by C2003. We make use of photometric methods to compute
effective temperatures and bolometric corrections, and we
require, therefore, good estimates of the interstellar
extinction. We adopted the $M_{\rm V}$ and E($b-y$) values
directly from ATT, except for HD~6833, for which
ATT did not derive such parameters, and HDE~232078, which ATT did
not study. Both stars lie at low Galactic latitude ($b$ = $-8.0$
and $-2.3$, respectively),
and are quite distant, so reddening is a special problem.

Table~1 summarizes the photometry
employed in our work. 
The $R-I$ photometry is on the Cousins system.
Thanks to the 2MASS
Point Source Catalog 
(Skrutskie et al.\ 2006)\footnotemark \footnotetext{The 2MASS photometry
discussed here were produced by the Two Micron All Sky Survey, which
was a joint project of the University of Massachusetts and the Infrared
Processing and Analysis Center/California Institute of Technology,
funded by the National Aeronautics and Space Administration and
the National Science Foundation.}, we were able to find $V-K$
photometry for all of our new program stars, although we transformed the 2MASS $K$
magnitudes to the ``CIT" system following the prescription given in
the Explanatory Supplement to the 2MASS Second Incremental Data
Release, followed by a transformation to the ``TCS" systems
as described by Alonso et al.\ (1994). 

Temperature estimates were obtained using the color-$T_{\rm eff}$
relations derived by Alonso, Arribas, \& Martinez Roger (1999, 2001), based on
Infrared Flux Method determinations.
The published relations employ the ``TCS" system $V-K$ colors, and Johnson
$R-I$ colors, so we transformed the ($R-I$)$_{\rm C}$ values
of Table~1 into the Johnson system, using the relations
given by Fernie (1983). When more than one temperature
estimate was available, as was most often the case, we employed
a simple mean.
The average rms scatter when three or more
temperatures estimates are available is 45~K.

We assumed stellar masses
of 0.8\msun, appropriate for stars at the
main sequence turn-off in globular clusters. 
Calculation of gravities then followed from 
the $M_{V}$ values, transformed to $M_{\rm bol}$ 
after addition of the bolometric
correction (Alonso et al.\ 1999).
Table~2 summarizes the resultant atmospheric parameters
for our program stars. 

\subsection{Special Cases}

To estimate the interstellar extinction to HD~6833, 
we exploited the spectroscopic study
by Fulbright (2000), who derived $T_{\rm eff}$ = 4450~K,
log~$g$ = 1.4, and [Fe/H] = $-1.04$. The star has
$V$ = 6.75 and $B-V$ = 1.17, and its metallicity is
a near-perfect match for that of NGC~6723 ([Fe/H] = $-1.03$
Kraft \& Ivans 2003). To obtain the same effective temperature
from the Alonso et al.\ (1999) calibrations requires E($B-V$) = 0.11
[E($b-y$) = 0.076], so ($B-V$)$_{0}$ = 1.06. The available
color-magnitude diagram of the cluster reveals that
M$_{\rm V} \approx\ -0.4$ at that color, which we adopt. Note that
as a result, our photometric estimate of log~$g$ is 1.6,
quite similar to that obtained by Fulbright (2000), who
derived a value of 1.4.

In the case of HDE~232078, we relied on the analysis
by Burris et al.\ (2000). They determined the temperature
and metallicity using spectroscopic methods, but employed
photometry to estimate the gravity, finding
$T_{\rm eff} = 4000$, log~$g$ = +0.3, and [Fe/H] = $-1.54$. These values enable us
to determine the interstellar extinction, under the assumption
that the Alonso et al.\ (1999) color-temperature calibration
should yield the same temperature if we have estimated
the reddening correctly. The globular cluster M3 has 
nearly identical metallicity ([Fe/H] = $-1.47$; Kraft et al.\ 1992),
and a very low reddening [E($B-V$) = 0.01 mag],
so in principle we only need to use the Alonso et al.\ (1999)
relations between $V-K$ and $T_{\rm eff}$ to estimate temperatures
of its red giants. Then we simply identify ($V-K$)$_{0}$ for
stars with the same effective temperature as HDE~232078. 
Using available unpublished M3 $BV$ photometry and 2MASS $K$
magnitudes, we obtain E($B-V$) = 0.56 mag for HDE~232078,
as well as $M_{\rm V}$ = $-2.15$. Our photometric
estimate for the gravity is log~$g$ = +0.6, somewhat higher
than that estimated by Burris et al.\ (2000).

\subsection{Tests of Distance, Gravity, \& [Fe/H] Estimations}

We rely on the photometric estimation of $M_{\rm V}$ values
and bolometric corrections to estimate the gravities of our
program stars,
so we must ask how accurate such estimates are, especially
given the relatively large distances to our program stars.
We have two methods available: trigonometric parallaxes from
HIPPARCOS, and spectroscopic gravities derived during the
course of metallicity determinations.

\begin{figure}
\epsscale{0.80}
\plotone{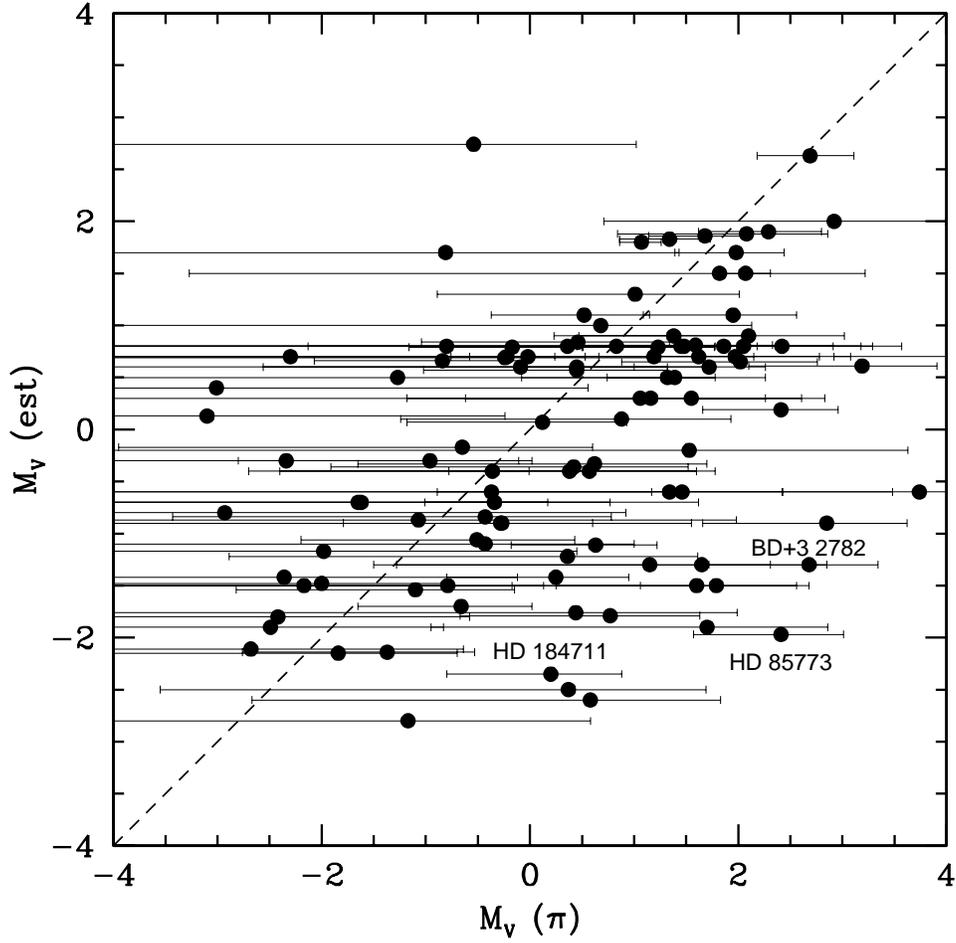}
\caption{A comparison of our estimated $M_{\rm V}$ values with
those derived using HIPPARCOS parallaxes. Stars with negative
parallaxes have been 
omitted, and were stars with $M_{\rm V} \leq\ -4.0$. Three stars
are flagged and discussed in the text.\label{fig1}}
\end{figure}

Figure~\ref{fig1} plots $M_{\rm V}$ values estimated by ATT
and, in the special cases here and in C2003, by us, against
those derived directly from trigonometric parallaxes. The 1$\sigma$ error
bars are almost all very large, since typical parallaxes are
only one to two milli-arcseconds, and the uncertainties are
often comparable to or even larger than the parallaxes, so
that the formal error encompasses physically unrealistic
negative parallaxes.
We have not plotted the 16 stars with negative parallaxes, nor
the 9 stars whose resultant $M_{\rm V}$ values are brigher
than $-4.0$ mag since those stars are more luminous than the
red giant branch tip stars in globular clusters. All 9
stars have very large errors in $M_{\rm V}$, the smallest
value being 4.85 mag. For the remaining stars whose error bars result in
negative parallaxes, we have applied very large error bars so
that they run off the left edge of Figure~\ref{fig1}. We acknowledge that
the Figure is not particularly helpful, but does merit
some scrutiny. For example, there are a number of stars for
which the parallaxes predict much fainter absolute magnitudes
and, therefore, much higher gravities. We have identified
three such stars in particular: BD+3~2782 ($M_{\rm V,adopted}$ = $-1.3$;
$M_{\rm V,\pi}$ = $+2.68^{+0.66}_{-0.96}$); HD~85773
($M_{\rm V,adopted}$ = $-1.97$; $M_{\rm V,\pi}$ = $+2.41^{+0.60}_{-0.84}$);
and HD~184711 ($M_{\rm V,adopted}$ = $-2.35$; $M_{\rm V,\pi}$ = 
$+0.20^{+0.68}_{-1.00}$). For these three stars, Gratton et al.\ (1996)
employed ionization balance to estimate gravities from
high-resolution, high-S/N spectra.
Our photometry-based gravities for these three stars are, respectively,
log~$g$ = +1.3, +0.90, and +0.59, consistent with high luminosity.
Gratton et al.\ (1996) obtained +1.28, +0.48, and +0.15, even
lower gravities, implying even higher luminosities
for HD~85773
and HD~184711 than we have estimated. We believe the few disagreements
between our derived $M_{\rm V}$ values and those obtained from
parallaxes reflect the uncertainties at the limit of
the Hipparcos mission capabilities.

\begin{figure}
\epsscale{0.80}
\plotone{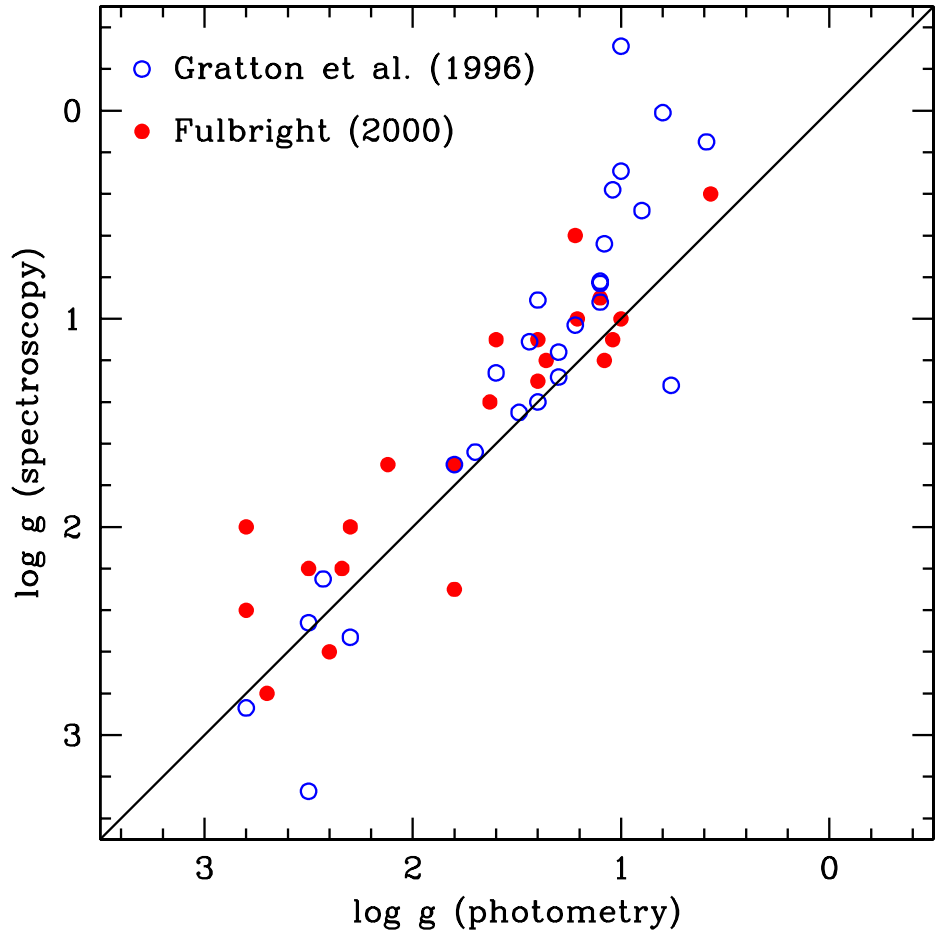}
\caption{A comparison of the gravities derived from
photometry with those obtained
spectroscopically by Gratton et al.\ (1996) and Fulbright (2000).
\label{fig2}}
\end{figure}

Figure~\ref{fig2} continues the comparisons of spectroscopic and
photometric gravities, relying on Gratton et al.\ (1996) as well
as Fulbright (2000). The Figure reveals that 
the gravities derived by Gratton et al.\ (1996) are lower
than the photometric estimates for almost all of the lowest gravity,
highest luminosity stars. Agreement with Fulbright (2000), however,
is much better. A least squares fit results in a slope of
almost unity (0.92) and modest offset (0.04 dex), and,
in particular, a small scatter ($\pm 0.27$~dex). That level
of scatter is very satisfying given the difficulties of estimating
$M_{\rm V}$ and, similarly, in determining spectroscopic gravities.

\begin{figure}
\epsscale{0.80}
\plotone{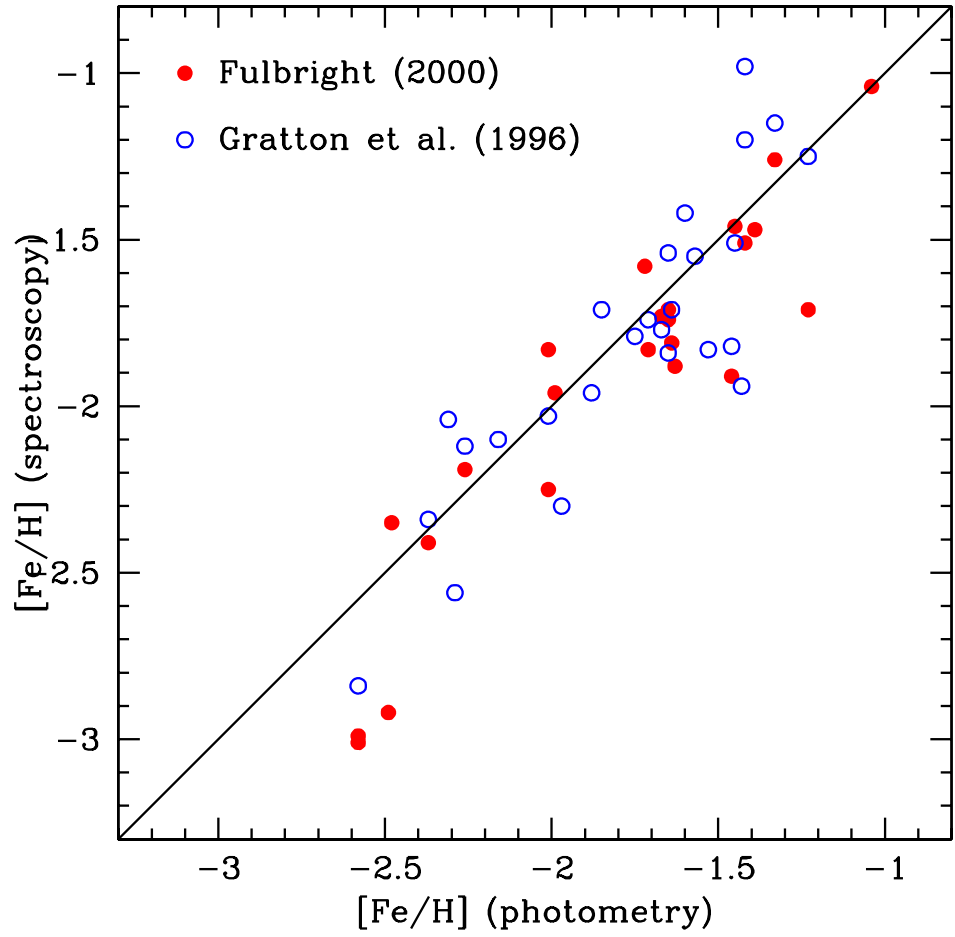}
\caption{A comparsison of our adopted photometric
metallicities with those obtained
spectroscopically by Gratton et al.\ (1996) and Fulbright (2000).
\label{fig3}}
\end{figure}

Figure~\ref{fig3} shows how our adopted [Fe/H] values, based primarily
on the Str\"{o}mgren photometry calibrations from ATT, compare with
results obtained from classical fine analyses of high-S/N, high-resolution
spectra of Gratton et al.\ (1996) and Fulbright (2002). The mean offset
in [Fe/H]
between our adopted values and the 26 stars in common with
Gratton et al.\ (1996) for our entire
survey (Carney et al.\ 2003 and this paper) is $+0.03 \pm 0.04$ dex,
with $\sigma = 0.22$ dex, in the sense our results minus those of
Gratton et al. For Fulbright (2000), the offset for 23
stars in common is $+0.12 \pm 0.04$ dex, $\sigma = 0.20$ dex.
Both of these results are comparable to the mean difference
between the 12 stars in our program that are common to both
Gratton et al.\ (1996) and Fulbright (2000): 
$<$[Fe/H]$_{\rm Gratton}$ - [Fe/H]$_{\rm Fulbright}>$ = $+0.14 \pm 0.06$,
$\sigma = 0.19$ dex.

\begin{figure}
\epsscale{0.80}
\plotone{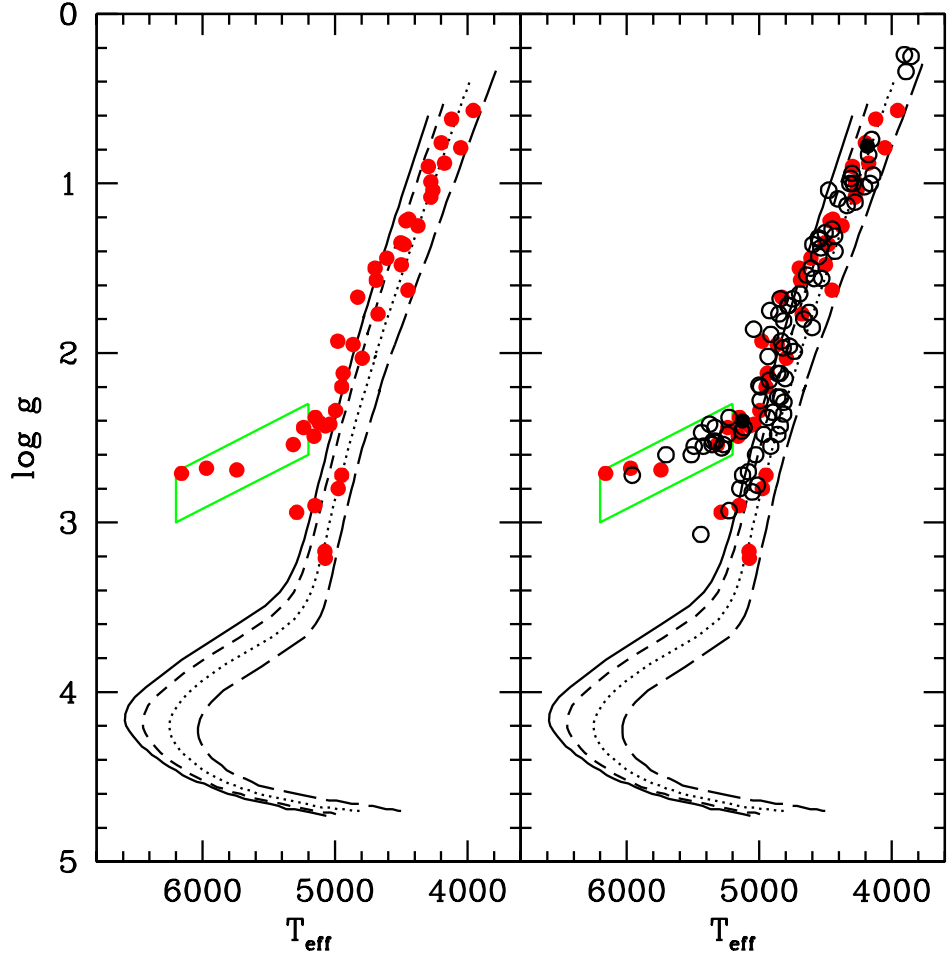}
\caption{A comparison between the gravities and temperatures
estimated for our new program stars (left panel) and
the combined samples from this paper and C2003 (right panel).
The model isochrones are taken from
Straniero \& Chieffi (1991), with
Z = 0.0001, 0.0003, 0.001, and 0.003.
The more metal-poor isochrones lie at higher temperatures.
The adopted age in all cases is 14 Gyrs.
Open circles denote the C2003 program stars. The parallelograms
enclose the red horizontal branch stars.\label{fig4}}
\end{figure}

Figures~\ref{fig1} through \ref{fig3} give us confidence that
our gravity, luminosity, and metallicity estimations are, in general, quite good.
Figure~\ref{fig4} plots the derived temperatures and gravities
for our program stars vs.\ the 14 Gyr model isochrones
of Straniero \& Chieffi (1991). The two panels show
the stars discussed in this paper, and the combined
results from this paper plus those stars studied by C2003.
There is excellent agreement between the data points
and the isochrones on the RGB, suggesting again that our
derived atmospheric parameters are quite plausible.

The reader may have noted (as did the referee) 
that the match to the isochrones
in Figure~\ref{fig4} is not perfect, with stars whose
gravities are only slightly lower than those of
horizontal branch hovering nearer
the hotter, more metal-poor isochrones than are most of the
lowest-gravity stars. There are several potential
contributing factors at work. For one, the two
sets of stars have slightly different metallicities.
The 25 stars with log~$g \leq\ 1.2$ 
have $<$[Fe/H]$>$ = $-1.86 \pm 0.06$ ($\sigma$ = 0.31),
while the 35 stars with 1.6 $\leq$ log~$g$ $\leq\ 2.3$
have $<$[Fe/H]$>$ = $-2.14 \pm 0.10$ ($\sigma$ = 0.57).
So a modest shift to more metal-poor isochrones is
expected. But there are subtle systematic effects that
may also contribute. For example, the lower gravity stars
are more luminous and intrinsically rarer, and, hence,
more distant. Reddenings may be more susceptible to
systematic errors. And the color-temperature transformations
may not be as well calibrated in one gravity domain compared
to another. We judge the overall agreement between our
estimated gravities and temperatures compared to the
model isochrones to satisfy the basic test that, on balance,
we have reliable stellar atmospheric parameters for our stars.

\subsection{Selection of Synthetic Spectrum Templates}

We used these derived atmospheric parameters to
select the optimum synthetic spectrum 
to derive both the radial velocities and the
line broadening. The average
correlation value was computed for all the observed spectra of each star
using all available rotationally-broadened templates with the
adopted values of temperature, gravity and metallicity.
The template with rotational broadening that 
yielded the highest average correlation was chosen for final
use in all the radial velocity measures.
The final column of Table~2 includes $T_{\rm eff}$, log~$g$, [Fe/H],
and $V_{\rm broad}$ for the synthetic spectrum employed.

\section{RADIAL VELOCITIES}

Table~3 provides the individual heliocentric radial
velocity results derived for each star
using the tool {\bf rvsao} (Kurtz \& Mink 1998) running inside the
IRAF\footnotemark \footnotetext{IRAF (Image Reduction and Analysis
Facility) is distributed by the National Optical Astronomy
Observatories, which are operated by the Association of Universities
for Research in Astronomy, Inc., under contract with the National
Science Foundation.} environment. The electronic version of the
table includes {\em all} the radial velocities from C2003
as well as those obtained from this survey. We do this primarily
for the convenience of the reader, and to assure ourselves that
all the data have been treated consistently. 
The results quoted in
Table~3 reflect only the internal errors. The reader should also
be aware that the velocities are on the ``native CfA"
system defined by our grid of synthetic spectra and
nightly observations of the dawn and dusk sky. 
To transfer these values to an absolute system defined by
observations of minor planets, 0.139 \kms\ should be added
to these values (Stefanik 1999).
Gravitational redshifts have also been neglected.

Table~4 summarizes the new results
for the 11 stars from C2003 which we have continued
to observe, while Table~5 does the same for the
new program stars. In both cases, we list the 
number of observations, the span of the observations in days, the mean
radial velocity, and the uncertainty of the mean velocity. Note that
for the binary stars, the mean radial velocity is not as appropriate
as the systemic velocity that emerges from the orbital solution. For
stars with orbital solutions, we therefore list here the systemic
velocity and its uncertainty. We also list the measured rms external
error, E, and I, the mean of the internal errors, $\sigma_{\rm int}$, 
of the velocity measurements
(see Kurtz \& Mink 1998). We also show the ratio, E/I, since large values of E/I ($> 1.5$)
are suggestive of radial-velocity variability. 

As discussed by C2003,
another powerful indicator of radial velocity variability is the
probability, P($\chi^{2}$), that the $\chi^{2}$ value could be larger
than observed due to Gaussian errors for a star that actually has
constant velocity.  We employ the internal error estimate, $\sigma_{\rm int}$,
for each of $n$ exposures when calculating
$\chi^2$:
\begin{equation}
\label{eq:chisquared}
\chi^2 = \sum_{i=1}^n (\frac{x_i-<x>}{\sigma_{i,\rm int}})^2 .
\end{equation}
The values of $\sigma_{\rm i,int}$ are computed by {\bf rvsao}.
If these internal error estimates are correct and the
actual errors are Gaussian, then the distribution of P($\chi^{2}$) should
be constant from 0 to 1, except for the velocity variables 
occupying the region near
zero probability. However, we found that use of the internal error
estimates led to a distribution of 
P($\chi^{2}$) values that was not flat
for P($\chi^{2}$)$ > 0.05$. Since a flat distribution is expected,
we found it necessary to add in quadrature a
``floor error" of 0.25 \kms\ to each internal error estimate.  This floor
error compensates for various systematic errors, such as shortcomings in
the run-to-run zero-point corrections, which are not included in the
internal error estimates.
The P($\chi^{2}$) values, derived with the additional
``floor error", are listed in Tables~4 and 5. All stars which we
have found to be binaries had P($\chi^{2}$) $< 10^{-6}$. 

\section{VELOCITY VARIABLES}

\subsection{Identification of Velocity Variables}

\begin{figure}
\epsscale{0.80}
\plotone{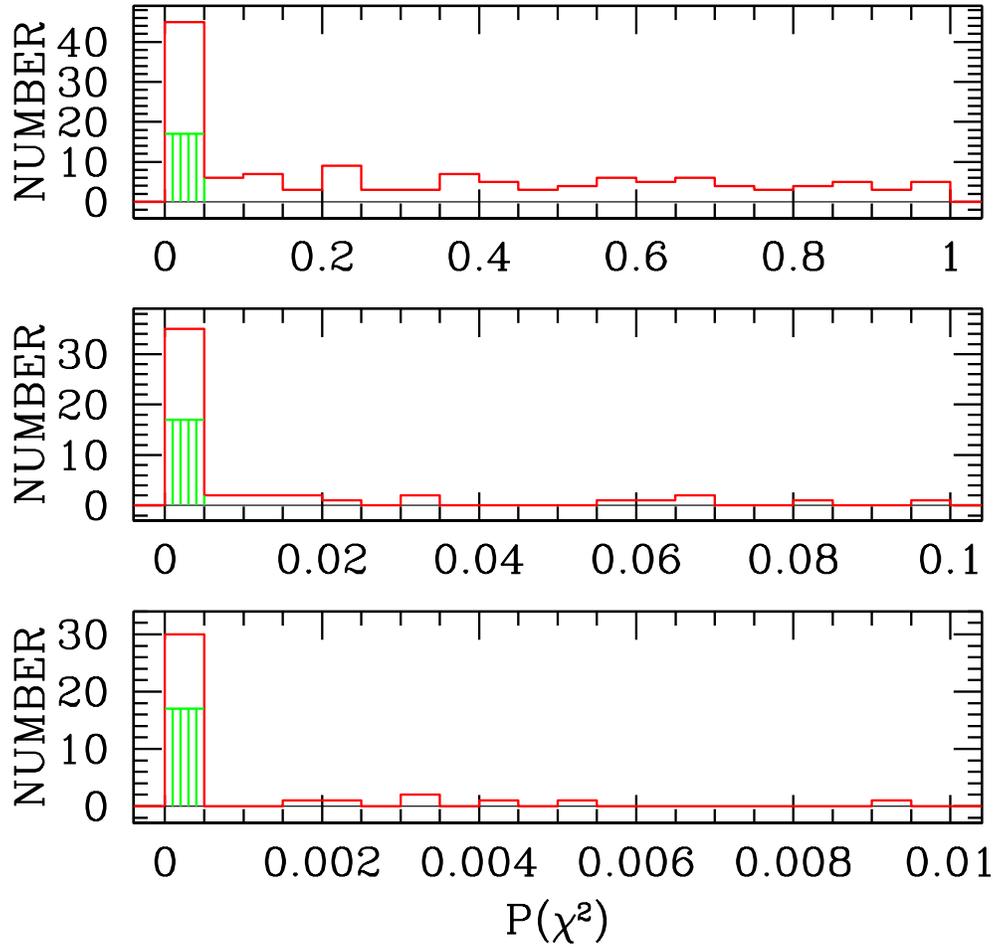}
\caption{The distribution of the probabilities of $\chi^{2}$
for our program stars, excluding the three spectroscopic binaries
for which orbital solutions are not yet available (HD~4306,
BD$-1$~2582, and HD~108317).
There is an excess of small values, indicating a substantial
number of stars
with variable velocities. Vertical shading indicates stars found
to be spectroscopic binaries. \label{fig5}} 
\end{figure}

Figure~\ref{fig5} shows the distribution of P($\chi^{2}$) values
for 135 of the 138 stars in C2003 and this paper. For stars with
spectroscopic binary orbital solutions, we use the values
derived by C2003 from the orbital solutions rather than
the values based on the velocity data alone.
We have {\em not} plotted here the three
stars that are clearly binary systems but which do not yet
have orbital solutions,
HD~4306, BD$-1$~2582 and HD~108317. 
As expected,
the probability distribution is flat, except for an
excess at the lowest values, which we magnify in the
bottom two panels. The shaded areas reveal stars known
to be spectroscopic binaries, while the unshaded region
in the lowest bin shows the stars that are mostly,
if not entirely, velocity variables and whose explanation
must involve pulsation or velocity jitter.

\subsection{Stars Studied Previously}

We begin with the data obtained for the 11 stars discussed already
by C2003 for which we have obtained additional observations.
BD+3~2782 shows no sign of any velocity variability, as before.

Three stars, HD~97, HD~3179, and HD~213467, continue to have
low P($\chi^{2}$) values, but other than one or two deviant
velocities, appear to have stable velocities, 
and it is therefore hard to
ascribe the apparent variability as due to either orbital motion
or velocity jitter. 

Four stars are spectroscopic binaries
with orbital solutions presented in C2003 (HD~6755, HD~27928,
HD+18~2796, and HD+1~3070). None of the orbital solutions
are affected significantly, except that the period uncertainties for HD~6755,
BD+18~2796, and BD+1~3070 improve from 
$\pm24$ days to $\pm 4.6$ days, 
from $\pm 71$ days to $\pm 54$ days, and
from $\pm 12$ days to $\pm 7$ days, respectively.

\begin{figure}
\epsscale{0.80}
\plotone{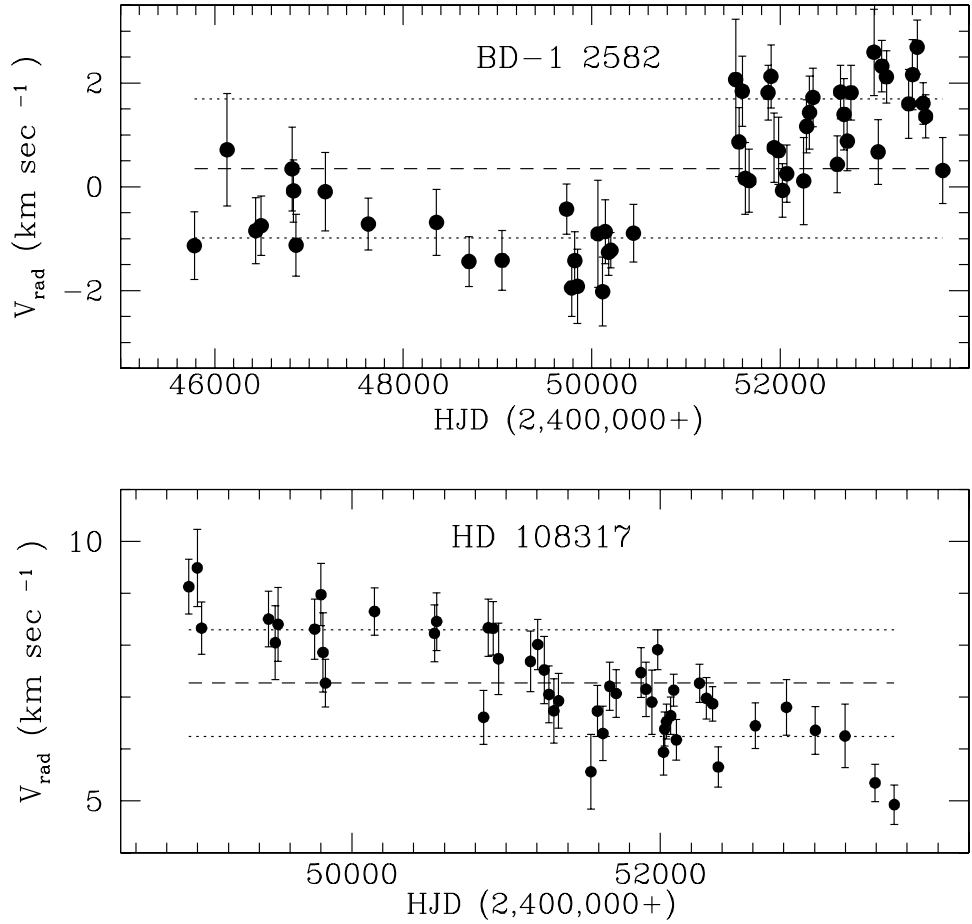}
\caption{The radial velocity histories of 
two binary stars from C2003, for which we have
obtained additional observations, but whose orbital solutions 
are as yet incomplete:
(a) BD$-1$~2582, and (b) HD~108317. The mean velocities as
well as $\pm 1\sigma_{\rm int}$ are also shown as horizontal lines.
\label{fig6}}
\end{figure}

The two recognized spectroscopic binaries whose orbital solutions
could not be determined by C2003, BD$-1$~2582 and HD~108317, remain
resistant to solution. Figure~\ref{fig6} shows the current status
of our observations. BD$-1$~2582 now has 52 velocities
spanning 7938 days (21.7 years), compared to the results
presented in C2003 (37 velocities over 6562 days), while
HD~108317 now has time coverage increased from 3432 days
to 4573 days (12.5 years) and the number of velocities
has increased from 42 to 48. 

HD~121261 remains hard to classify. We have increased the
time coverage from 5142 days to 7684 days (21.0 years), and the
number of velocities from 12 to 15. Unfortunately, there is
still no suggestion of a periodicity, and the velocity scatter
remains modest (0.85 \kms). The star is judged to be relatively
luminous, $M_{\rm V} \approx\ -1.5$, so velocity jitter
could also be responsible.

\subsection{New Velocity Variables}

\begin{figure}
\epsscale{0.80}
\plotone{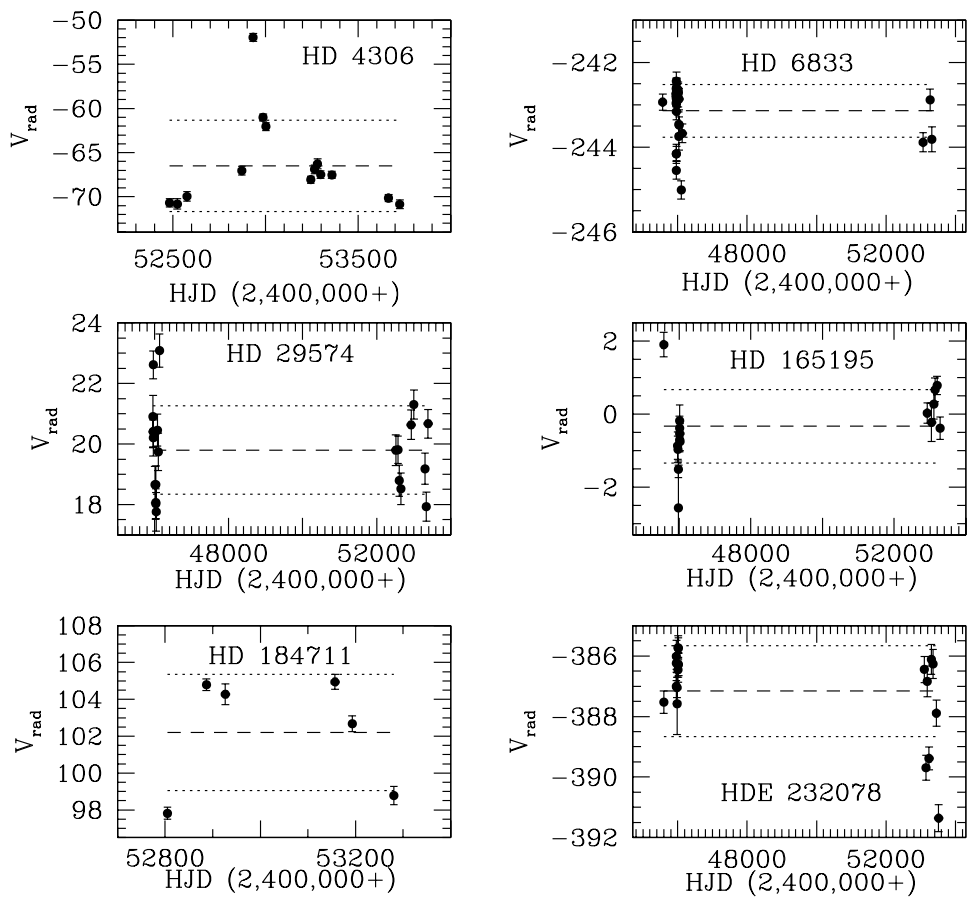}
\caption{The velocity histories, in \kms, of 
the six stars in our new study with P($\chi^{2}$) $\leq\ 0.001$.
(a) HD~4306; (b) HD~6833; (c) HD~27295; (d) HD~165195;
(e) HD~184711; and (f) HDE~232078. The mean velocities as
well as $\pm 1\sigma_{\rm int}$ are also shown as horizontal lines.
\label{fig7}}
\end{figure}

Following Figure~\ref{fig5}, we discuss only those stars 
in our new sample with P($\chi^{2}$) values 
lower than 0.001. Figure~\ref{fig7} shows the velocity histories
for all six stars. 

HD~4306 appears to be a spectroscopic binary with an
eccentric orbit. HD~184711 is a less compelling case
for being a spectroscopic binary, but more observations
are needed to confirm this conjecture. The star is
very luminous ($M_{\rm V} \approx\ -2.3$), so the
variability could as easily be ascribed to velocity jitter.

HD~6833, HD~29574, BD+1~2916, HD~165195, and HDE~232078
all have good velocity coverage and 14 or more measured
velocities, and all have very low P($\chi^{2}$) values
($< 10^{-6}$). None show any signs of periodicity in
their velocities. All of them, except HD~6833, are
judged to be luminous, with $M_{\rm V} \leq\ -1.7$,
so velocity jitter seems to be the best explanation.

\subsection{Velocity Jitter Revisited}

\begin{figure}
\epsscale{0.80}
\plotone{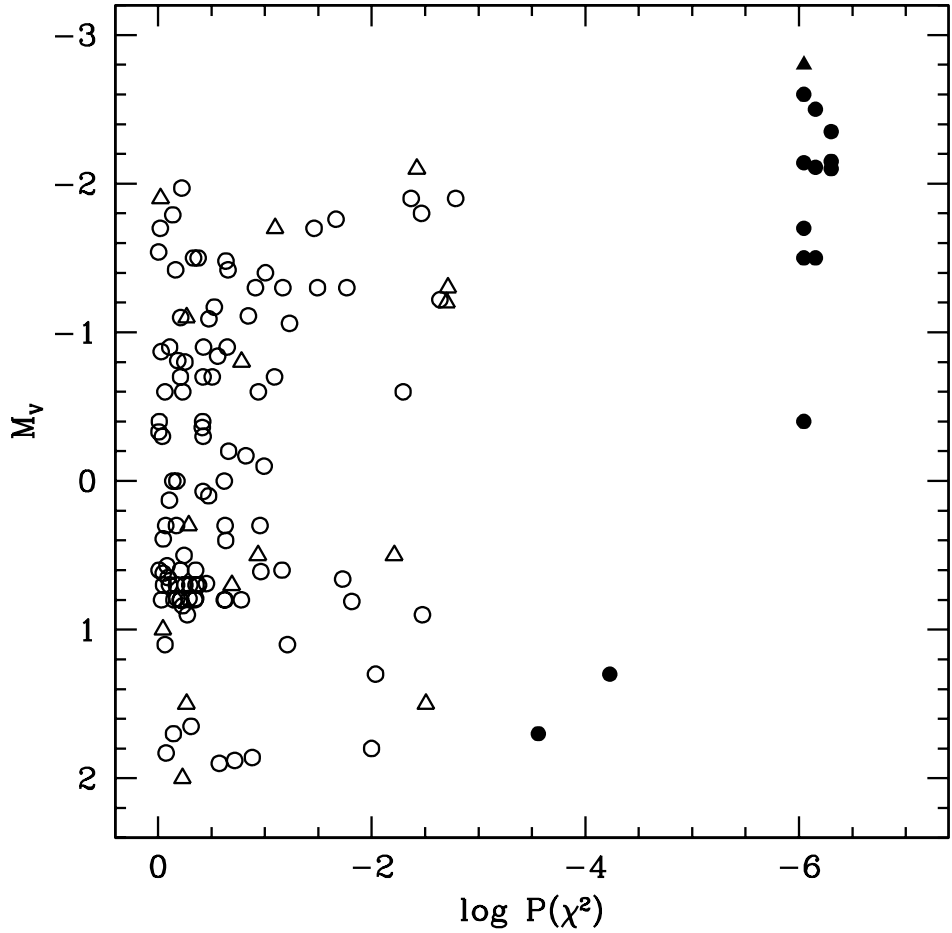}
\caption{The logarithms of the P($\chi^{2}$)
values are plotted against the estimated
absolute magnitudes. 
Non-binary stars with P($\chi^{2}$) $\leq$\
0.001 are plotted as filled circles and are assumed to be displaying
velocity jitter. 
All other non-binary stars are plotted as open circles.
The spectroscopic
binary stars' P($\chi^{2}$) values have been computed using the
residuals from the orbital solutions. 
The binary stars are represented by triangles.\label{fig8}}
\end{figure}

Figure~\ref{fig8} is an update of Figure~8 in C2003, with
the stars from Table~5 added. Binary stars are shown as
triangles; non-binary stars as circles.
All stars with P($\chi^{2}$) 
$\leq\ 0.001$ are plotted as filled triangles or circles.
We have not included HD~4306 because it is uncertain
if the very low P($\chi^{2}$) value is due to its orbital
motion, jitter, or both. Its low luminosity ($M_{\rm V} \approx\ +0.2$)
and its velocity vs.\ time behavior (Figure~\ref{fig7})
suggest that only orbital motion is at work, but until we have
a reliable orbital solution, we cannot say for certain.

The most striking feature of Figure~\ref{fig8} is that the lowest
P($\chi^{2}$) values, $\leq\ 10^{-6}$, are restricted to the most luminous stars,
those with $M_{\rm V} \leq\ -1.4$. The second striking
feature is that not all stars with such luminosities manifest
velocity jitter. We have included binary stars with orbital solutions,
and their 
P($\chi^{2}$) were recomputed following
application of the orbital solution (see the discussion
in C2003). Thus the P($\chi^{2}$) values are a measure of non-orbital sources
of velocity variability. We find that only 3 of the 14 stars
with estimated $-2 < M_{\rm V} \leq\ -1.4$ show P($\chi^{2}$)
values of less than $10^{-6}$. But {\em all seven} of the stars with
estimated $M_{\rm V}$ values of less than $-2.0$ do show
jitter. Our estimated $M_{\rm V}$ values do not have
sufficient precision to ascertain if velocity jitter
``turns on" at a particular luminosity or gravity, but
that is a reasonable speculation.

Finally, we note the remaining challenge presented by
the three lower luminosity/higher gravity stars
with P($\chi^{2}$) $< 10^{-3}$. As noted
previously, HD~97 and HD~213467 ($M_{\rm V}$ = +1.4 and +1.7,
respectively) may have low P($\chi^{2}$) values due to
one or two poor radial velocity measures. HD~6833, however,
does appear to be a legitimate low-luminosity star with
variable velocity that has not, as yet, been coupled with
orbital motion. Our estimated gravity, log~$g$ = $1.63$,
agrees well with that obtained spectroscopically by
Fulbright (2006), log~$g$ = 1.4. Perhaps additional radial velocity monitoring
would reveal the star to be a binary.

\section{LINE BROADENING}

\subsection{Measuring Line Broadening}

We have estimated the line broadening, $V_{\rm broad}$, following the
procedures described by C2003. 
For each star we selected
the synthetic spectrum for use as a template that
best matches the derived temperature, gravity, and metallicity parameters,
and then chose the
rotational broadening value
which produced the highest average correlation value. 
We interpolated the derived correlation values
for the two velocities adjacent to the preferred value
using a quadratic fit. The derived $V_{\rm broad}$ values
are listed in the final column of Tables~4 and 5. The typical
standard error in these values derived from the individual spectra
ranged from 0.5 to 2.0 \kms, so the mean values are well
determined, at least internally. The systematic effects
were also discussed by C2003. Basically,
changes of 250~K in temperature,
0.5 dex in gravity, or 0.5 dex in metallicity, individually led
to velocity sensitivities of 3.0, 3.2, and 3.3 \kms,
independent of the derived line broadening. 
The synthetic spectra were all
computed using a macroturbulent velocity of 3~\kms,
but low gravity stars may have higher turbulent
velocities (see Gray 1982; Gray \& Pallavicini 1989).
The line broadening values for more luminous stars
would be expected to be affected more strongly
by macroturbulence at the higher luminosities.
Values that are well below our instrumental resolution
are more vulnerable to these uncertain systematic effects.

The referee has inquired about observing conditions when
the star underfills the spectrograph's slit, which may
lead to a higher instrumental resolving power than we
have assumed. For spectra obtained in Massachusetts,
(identified by ``W" in
the detailed velocity summary), the seeing always overfilled
the slit by a factor of two or more. In Arizona
(identified by ``T" or ``M"), there
were occasional times when the slit was underfilled.
We compared line broadening results for those stars
with spectra obtained in both Arizona and Massachusetts
and found no discernible differences.

\subsection{Comparisons with Other Studies}

The first question we ask is whether our results are
in reasonable agreement with other studies, especially
those obtained using higher spectral resolution and
higher S/N. We have made comparisons with three such
studies.

We have fifteen stars in common with Behr (2003), with a
span in $V_{\rm broad}$ from about 3 to 21 \kms.
Behr's observations included higher resolving power,
60,000 (about 5.0 \kms), higher S/N (for the
stars in common, his cumulative S/N for the
stars ranged from 40 to 200, with 60 being a typical value),
and greater wavelength coverage (1000 to 1500 \AA).
Nonetheless, despite our limited wavelength
coverage, lower resolving power, and
lower S/N observations, a linear least squares bissector
analysis results in 
\begin{equation}
\label{eq:vrotbehr}
V_{\rm broad,CfA} = 1.10 \times V_{\rm broad,Behr} + 0.00.
\end{equation}
The scatter about this relation is only 1.2 \kms. Taking
the 15 stars in a star-to-star comparison shows a mean
difference in the values of only 0.6 \kms,
again, a scatter of 1.2 \kms.
(The sense is CfA results minus those of Behr 2003.)
If we restrict the comparison to only the 7 stars for which
our results suggest $V_{\rm rot}$~sin~$i$ $\leq\ 6$ \kms,
the mean offset is only 0.3 \kms, with $\sigma = 1.1$ \kms.

We compare the results from this paper
and from C2003 with the sample of 20 stars for which
we obtained spectra with very high resolving power (150,000,
or 2.0 \kms) and very high S/N (generally over 200)
using the Gecko spectrograph on the Canada-France-Hawaii
Telescope. Details may be found in another paper (Carney et al.\ 2007).
These spectra were analyzed by Fourier transform methods
(see Gray 1982), which result in $V_{\rm rot}$~sin~$i$ and
$\zeta_{\rm RT}$, a measure of the star's macroturbulence dispersion.
To derive a ``combined" measure of line broadening, $V_{\rm broad}$,
we treat both values as Gaussians, which is not strictly correct,
since the macroturbulent dispersion, $\zeta_{\rm RT}$,
does not behave as a Gaussian. Nonetheless, we define
\begin{equation}
\label{eq:vbroad}
V_{\rm broad,CFHT}=[(V_{\rm rot} \sin\ i)^{2} + f\zeta_{\rm RT}^{2}]^{1/2}.
\end{equation}
This formulation is due to Massarotti et al.\ (2007), and as in
their study, our
value of $f$ was determined empirically, such that the mean
offset between the $V_{\rm broad,CfA}$ and $V_{\rm broad,CFHT}$ values
was minimized. We found the best agreement when $f$ is 0.95, at least
for the current sample of metal-poor giants. For the seven stars with 
$V_{\rm broad,CfA} < 8.5$ \kms, the mean offset is $0.0 \pm 0.4$ \kms,
with $\sigma = 0.9$ \kms. For the eight stars with 
$V_{\rm broad,CfA} \geq\ 8.5$ \kms, the mean offset is $0.0 \pm 0.5$ \kms,
with $\sigma = 1.4$ \kms, and for all fifteen stars, the mean offset
is $0.0 \pm 0.3$ \kms, $\sigma = 1.1$ \kms.

\begin{figure}
\epsscale{0.80}
\plotone{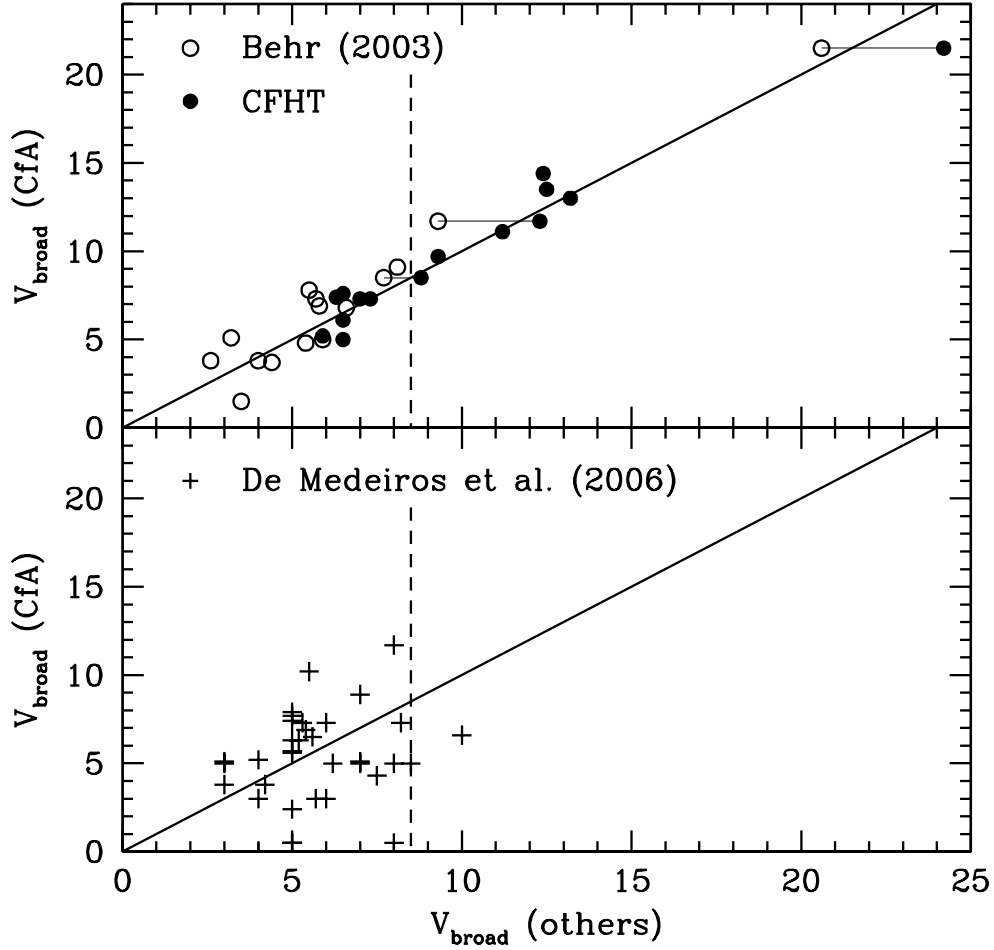}
\caption{A comparison of our derived
line broadening measures, in \kms, with those of Behr (2003),
de~Medeiros et al.\ (2006), and our own very high-resolution,
high-S/N spectra obtained using the Gecko spectrograph
at CFHT. The diagonal line is not a fit, but a unit slope.
The vertical dashed line represents the instrumental resolution
of the CfA equipment, about 8.5 \kms. Thin solid lines
connect the two measures obtained for three stars from Behr (2003)
and our CFHT spectra. \label{fig9}}
\end{figure}

The top panel of Figure~\ref{fig9} compares the results from
Behr (2003) and our CFHT program
with those of this paper and C2003. For
the CFHT sample, we have eliminated the five variable stars since
the temperatures and gravities derived from photometry may be
less reliable. For three stars, HD~25532, HD~184266, and HD~195636,
line broadening was determined by both Behr (2003) and from our CFHT
data, and we have drawn lines connecting the two sets of results.
Agreement is excellent, even well below our instrumental
resolution of 8.5 \kms. 

Another recent study is that of
de~Medeiros et al.\ (2006), who measured $V_{\rm rot}$~sin~$i$
(i.e., $V_{\rm broad}$)
values at a typical resolving power of 50,000 (6.0 \kms),
S/N $\approx\ 80$, and wavelength coverage spanning
$\lambda\lambda$3500-9200. Their work had 22 stars in common
with C2003, and with the results from the present paper,
there are 35 stars in common. The range in ``rotational
velocities" is more limited, despite the larger sample
size. For only three of the stars in common have we
obtained a $V_{\rm broad}$ value larger than
our resolving power, and the largest of those is 
only 11.7 \kms. The simplest comparison is then just
that from star-to-star matches. The average difference, in
the sense of our values minus those obtained by
de~Medeiros et al.\ (2006) is $-0.4$ \kms, with
$\sigma = 2.7$ \kms. If we compare only the 21 stars
for which we estimate $V_{\rm broad}$ $\leq\ 6$ \kms,
the mean difference is $-1.8$ \kms, with
$\sigma$ = 2.5 \kms. The bottom panel of Figure~\ref{fig9}
shows the comparison graphically. As the quantitative
comparisons found, the scatter appears to be larger
in the bottom panel than in the top panel, although
most of these comparisons are for stars whose 
line broadening values are comparable to or smaller
than either our instrumental resolution or that of
de~Medeiros et al.\ (2006). We note that de~Medeiros
et al.\ (2006) compared their results with those
of Behr (2003). Taking straight differences, we find
a mean difference, in the sense of de~Medeiros et al.\ (2006)
minus Behr (2003), of $+0.4 \pm 1.0$ \kms, with 
$\sigma = 2.8$ \kms, for the eight stars common to both
programs. The scatter somewhat larger than the value of 1.4 \kms\
that de~Medeiros et al.\ quoted for comparisons of
all the stars common to C2003, Behr (2003), and Peterson (1983).

We conclude that our results are in good agreement with all
of the above other studies, although perhaps somewhat better
with Behr (2003) than with de~Medeiros et al.\ (2006).
Considering that our spectra have much lower resolution,
much lower S/N, and much smaller wavelength coverage,
the power of synthetic spectrum templates is
apparent.

\section{ROTATION OR  MACROTURBULENCE?}

\begin{figure}
\epsscale{0.80}
\plotone{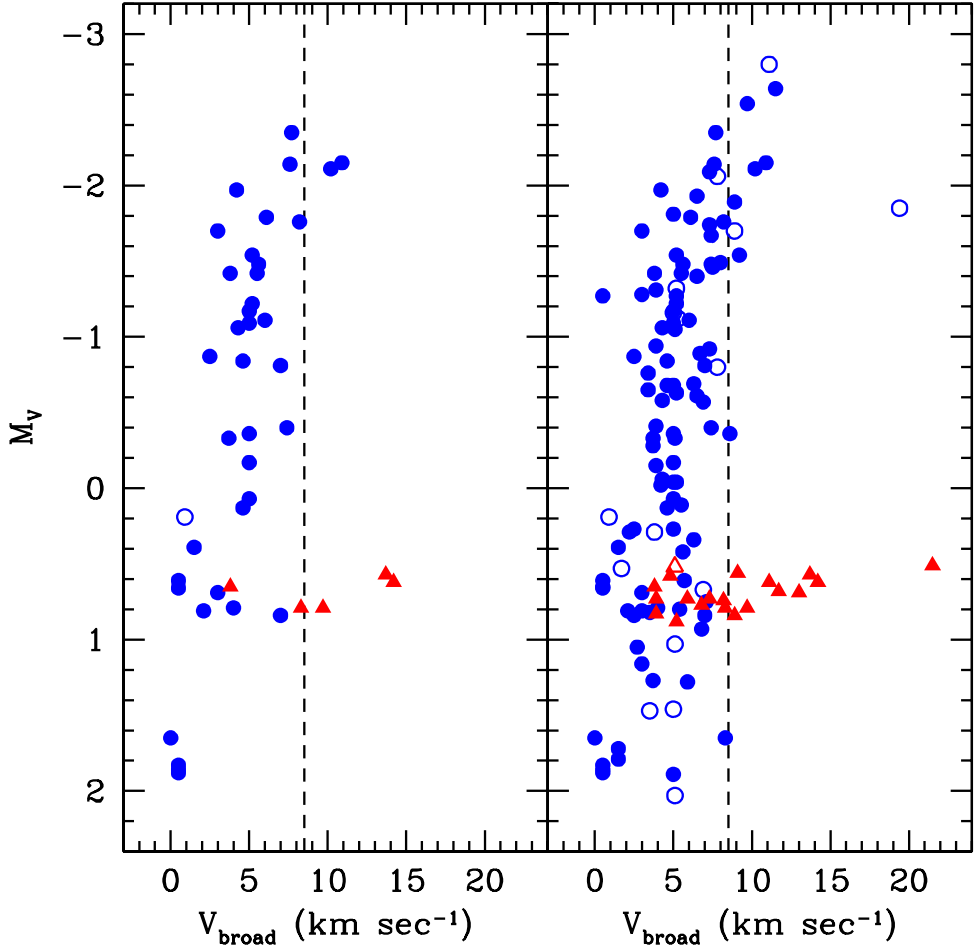}
\caption{The estimated line broadening 
in the spectra of the new program stars (left), and the
combined results from this program and C2003 (right).
Unlike the similar figure in C2003, we have plotted HD~3008 
and BD+22~2411 with their derived
$V_{\rm broad}$ values.
Open blue circles identify
the red giant binaries in the sample. Filled red triangles are used
to show the red horizontal branch stars enclosed within the
parallelogram of Figure~3. (The open red triangle is the
red horizontal branch binary star, HD~108317.)
The dashed line
indicates our instrumental resolution of 8.5~\kms.\label{fig10}}
\end{figure}

Figure~\ref{fig10} shows the line broadening for the stars
from Table~5 (left), and those stars plus the results
from C2003 (right). The 45 stars for which we have obtained
new results behave the same as the 91 stars studied by
C2003. Specifically, stars with $M_{\rm V} < -2$ show
systematically higher levels of line broadening than the
fainter stars. As noted earlier, C2003 speculated that while
some of the cases might be due to tidal locking in a binary
system\footnote{A prime example of this is CD$-$37~14010,
with $M_{\rm V}$ = $-1.9$, $V_{\rm broad}$ = 19.4 \kms,
and an orbital period of 65.55 days.}, some of the 
apparently elevated line broadening might
be due to increased rotation due to the absorption of a giant
planet that had an orbital separation of about one AU.
This idea was also consistent with the generally high levels
of line broadening seen in the RHB stars,
which are, of course, the direct descendents of some of the
RGB tip stars. The significant line broadening of many
of the RHB stars in our program is also apparent in
Figure~\ref{fig10}. 

Line broadening may also be due to
macroturbulence. C2003 dismissed this as a cause because
the levels of macroturbulence necessary to provide a total
line broadening of 9 to 10 \kms\ and higher seemed
improbable. Gray (1982) and Gray \& Pallavicini (1989) reported
macroturbulent velocities for K giants (luminosity class III),
and all values were smaller than 6.5 \kms. Even considering
luminosity classes II-III and II, Gray \& Toner (1986)
found macroturbulent velocities to be smaller than 8 \kms.
However, these results were derived from studies of metal-rich disk stars.
An obvious question is whether
lower-mass, older, more metal-poor halo giants
have elevated levels of macroturbulence compared to
disk giants. Carney et al.\ (2007) do not, in fact,
find significant differences in $\zeta_{\rm RT}$ between
metal-rich disk and metal-poor halo red giants.

\begin{figure}
\epsscale{0.80}
\plotone{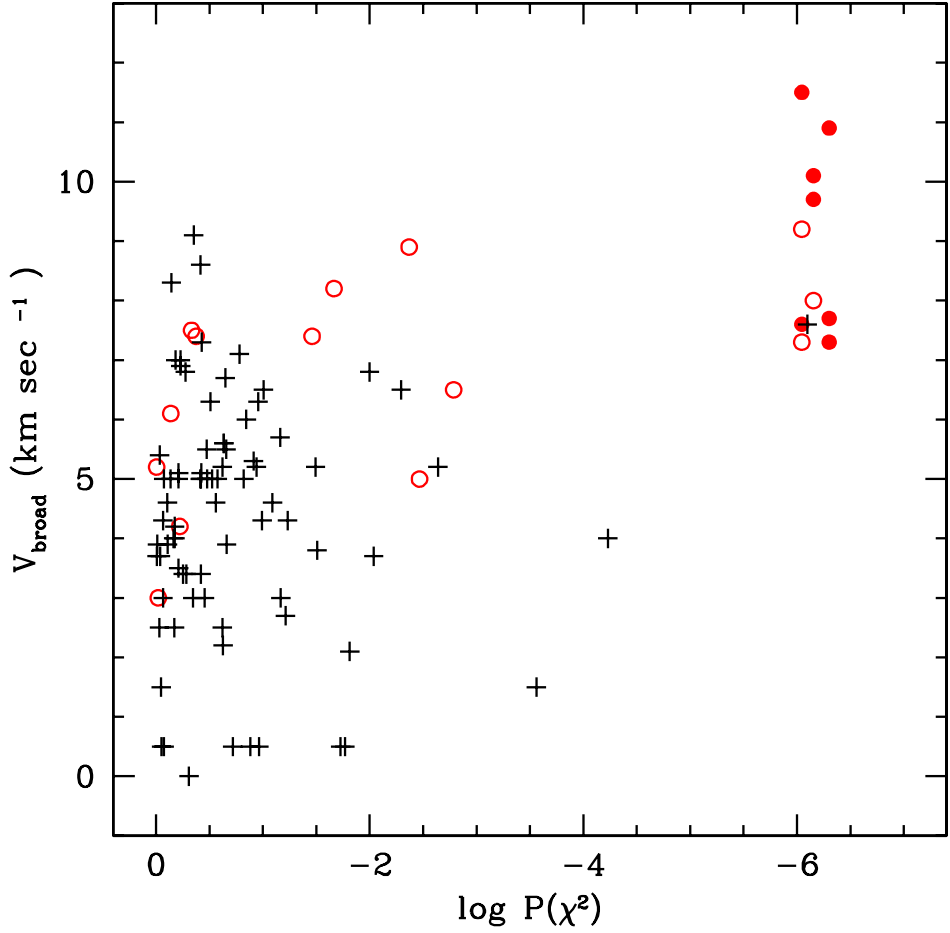}
\caption{The logarithm of the probability
of $\chi^{2}$ is compared with the measured line broadening,
$V_{\rm broad}$. Stars with $M_{\rm V} \leq\ -2.0$ are
shown as filled red circles, and those with
$-2.0$ $<$ $M_{\rm V} \leq\ -1.5$ are shown as
open red circles.\label{fig11}}
\end{figure}

What about velocity jitter?
Let us consider only the RGB stars and exclude
the known binary systems, where tidal interactions may
have contributed to the rotation and line broadening.
We have noted that velocity jitter begins to
appear at only the highest luminosities, $M_{\rm V} \leq\ -1.4$,
and especially for $M_{\rm V} \leq -2.0$. These are also
the luminosity levels for which line broadening measures
are highest. In Figure~\ref{fig11} we replace
luminosity with 
log~P($\chi^{2}$) and compare directly with $V_{\rm broad}$. Open
red circles are stars with $-2.0 < M_{\rm V} \leq\ -1.5$. 
Filled red circles are stars with $M_{\rm V} \leq\ -2.0$. As the Figure
shows, some of the luminous RGB stars have normal 
P($\chi^{2}$) and modest but typical line broadening. However,
a significant
number of the most luminous stars have very low values of P($\chi^{2}$), 
and they also tend
to have much higher than average values of $V_{\rm broad}$. 
In other words, it appears that there is a strong correlation
between velocity jitter and line broadening. This is not
a matter of ``velocity smearing" during individual observations.
The exposure times are far too short, typically a few minutes,
compared to the physical timescales for major atmospheric
changes in very large red giant stars (with timescales of
days to months).  This Figure suggests that velocity jitter
and line broadening share a common cause.

\begin{figure}
\epsscale{0.80}
\plotone{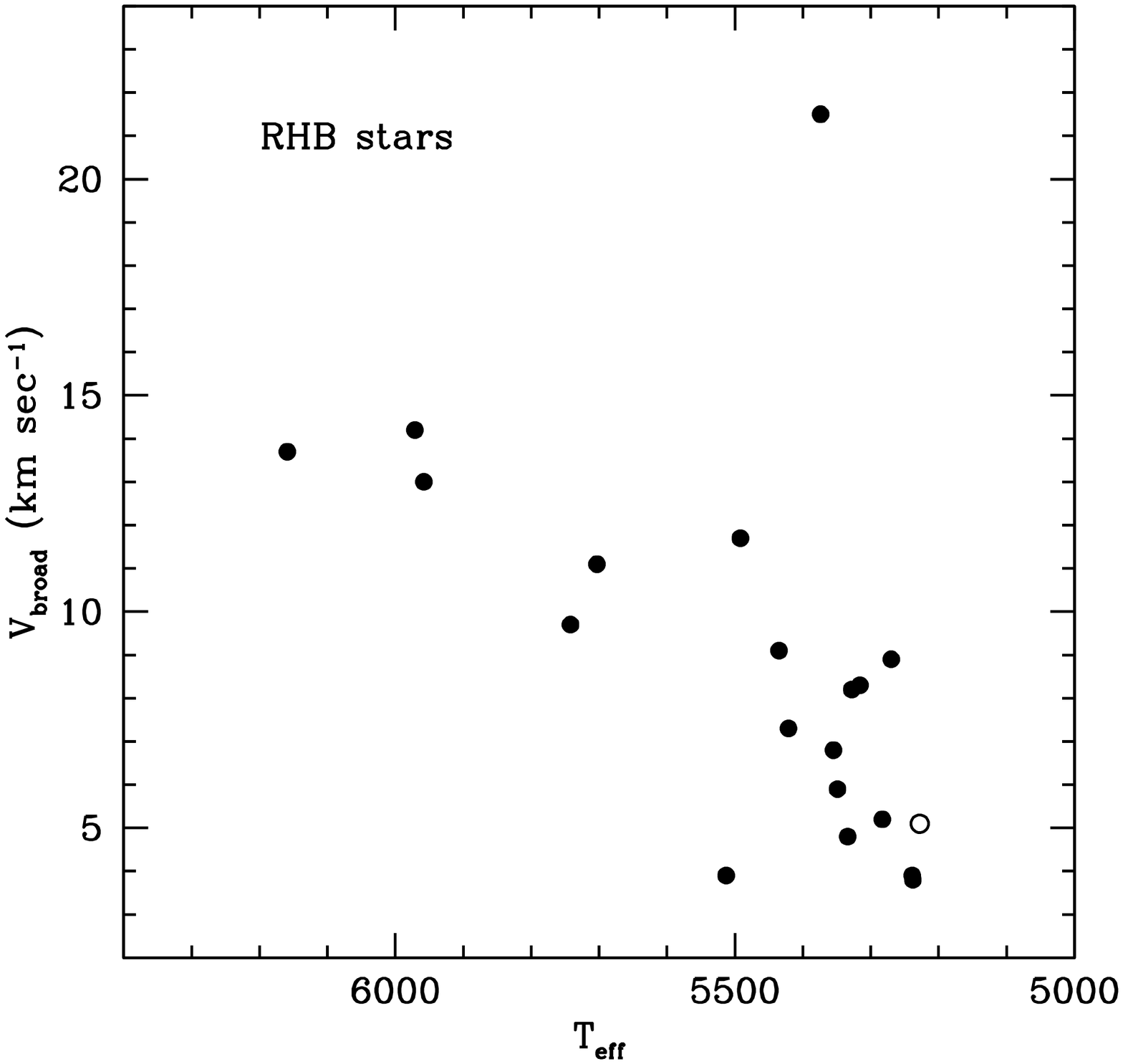}
\caption{The measured line broadening
is compared to the estimated effective temperature of the
red horizontal branch stars in this study and that of C2003.
Macroturbulence is expected to increase with temperature
at constant luminosity. The open circle is the 
long-period spectroscopic
binary HD~108317.\label{fig12}}
\end{figure}

How do we explain the significant line broadening measured
for the RHB stars? Rotation may play a significant role,
but we suspect the macroturbulence contributes to the broadening.
Gray (1982) first noted the
correlation between macroturbulent velocities and
effective temperature in giant stars,
with $\zeta_{\rm RT}$
rising from around 4 \kms\ for K2~III stars
to 7 \kms\ for G2~III stars. Gray \& Toner (1986) found similar
behavior for bright giants, luminosity classes II and II-III.
Figure~\ref{fig12} shows our measured line broadening for RHB
stars as a function of $T_{\rm eff}$. Qualitatively,
the trend matches the expected increase in $\zeta_{\rm RT}$
as a function of $T_{\rm eff}$. There are quantitative
difficulties, however, since the line broadening seen among
the RHB stars is much larger than the
macroturbulent velocities seen in metal-rich disk giants.
Rotation should certainly not be ruled out, either.
If the most luminous RGB stars have only very modest levels
of rotation, their smaller descendents should have significant
rotation.
Very high-resolution, high-S/N spectra of such stars will enable
us to explore the roles of rotation and macroturbulence 
in both RGB and RHB stars.

\section{CONCLUSIONS}

We have obtained 349 new radial velocities and line broadening measures
for 45 metal-poor RGB and RHB stars, as well
as 57 such measures for 11 of the stars we studied previously (C2003).
A comparison of our derived values for line broadening with results from
Behr (2003) and de~Medeiros et al.\ (2006) shows that our lower-resolution,
lower-S/N, and limited wavelength
coverage  spectra yield excellent results. We believe the good
agreement testifies to the power of
high-resolution synthetic spectra as
templates. We have identified one new spectroscopic binary, HD~4306,
and a possible second one, HD~184711, although we note that the latter's
radial velocity variability may be due to velocity jitter rather than
orbital motion.

We draw attention to
the observed correlation between variable
radial velocity (``jitter") and line broadening. 
The significant
line broadening seen in the metal-poor field RHB stars is
hard to explain: it may be a combination of ``spin up" when
a slowly rotating luminous RGB star settles on to the horizontal
branch, or temperature-dependent macroturbulence may be involved.
Very high-resolution and very high-S/N spectroscopy should
reveal the relative balance of rotation and macroturbulence
in RGB and RHB stars.

\acknowledgements
We thank R.\ Davis for his many years of service maintaining the
database for the CfA digital speedometers, and J.\ Caruso,
J.\ Zajac, P.\ Berlind, and G.\ Torres for making many of the
observations. 
BWC thanks the National Science Foundation for grants AST-9988156
and AST-0305431
to the University of North Carolina. JBL 
thanks the National Science Foundation for grants AST-9988247
and AST-0307340 to Bowling Green State University.
We especially acknowledge
the great utility of the 2MASS database, as
well as the SIMBAD database, maintained by the
CDS in Strasbourg, France.

\clearpage

\begin{deluxetable}{l r r r r r r r r}
\tablewidth{0pc}
\tablenum{1}
\footnotesize
\tablecaption{Photometry of Program Stars \label{tab1}}
\tablehead{
\colhead{Star} &
\colhead{E($b-y$)} &
\colhead{($b-y$)} &
\colhead{$V-R_{J}$} &
\colhead{Ref} &
\colhead{($R-I$)$_{C}$} &
\colhead{Ref} &
\colhead{$V-K$} &
\colhead{Ref}}
\startdata
HD 2665      & 0.049 & 0.551 & 0.729   & 1 & \nodata &   & 2.18 & 6 \\
HD 2796      & 0.005 & 0.542 & 0.718   & 1 & 0.478   & 3 & 2.28 & 9 \\
HD 4306      & 0.020 & 0.531 & 0.711   & 1 & 0.475   & 3 & 2.18 & 7 \\
HD 6268      & 0.012 & 0.599 & \nodata &   & \nodata &   & 2.42 & 9 \\
HD 6229      & 0.020 & 0.489 & \nodata &   & \nodata &   & 2.05 & 9 \\
HD 6833      & 0.076 & 0.753 & 0.947   & 1 & \nodata &   & 2.94 & 9 \\
HD 8724      & 0.029 & 0.683 & 0.887   & 1 & \nodata &   & 2.73 & 9 \\
HD 11582     & 0.005 & 0.464 & \nodata &   & 0.431   & 3 & 1.91 & 9 \\
HD 13979     & 0.000 & 0.503 & \nodata &   & \nodata &   & 2.11 & 9 \\
CD$-36$ 1052 & 0.011 & 0.342 & \nodata &   & 0.320   & 3 & 1.46 & 9 \\
HD 21022     & 0.000 & 0.651 & \nodata &   & 0.530   & 3 & 2.59 & 9 \\
CD$-24$ 1782 & 0.001 & 0.468 & \nodata &   & 0.421   & 3 & 2.03 & 9 \\
HD 25532     & 0.053 & 0.482 & 0.660   & 1 & \nodata &   & 1.95 & 7 \\
HD 26297     & 0.001 & 0.739 & 0.925   & 1 & 0.603   & 3 & 2.89 & 7 \\
BD+6 648     & 0.088 & 0.875 & 1.066   & 1 & 0.737   & 3 & 3.28 & 9 \\
HD 29574     & 0.036 & 0.036 & 1.159   & 1 & 0.732   & 3 & 3.57 & 9 \\
HD 33771     & 0.015 & 0.591 & \nodata &   & 0.512   & 3 & 2.46 & 9 \\
HD 41667     & 0.000 & 0.624 & \nodata &   & \nodata &   & 2.60 & 9 \\
HD 44007     & 0.067 & 0.560 & 0.754   & 1 & 0.494   & 3 & 2.385 & 7 \\
HD 74462     & 0.017 & 0.665 & 0.850   & 1 & \nodata &   & 2.675 & 9 \\
BD+54 1323   & 0.003 & 0.470 & 0.642   & 1 & \nodata &   & 1.965 & 7 \\
BD+58 1218   & 0.000 & 0.515 & 0.695   & 1 & 0.450   & 4 & 2.175 & 8 \\
HD 85773     & 0.015 & 0.788 & 0.945   & 2 & 0.587   & 2 & 2.91 & 9 \\
HD 88609     & 0.000 & 0.675 & 0.852   & 1 & 0.570   & 4 & 2.59 & 8 \\
HD 101063    & 0.027 & 0.499 & \nodata &   & 0.459   & 3 & 2.11 & 9 \\
BD+52 1601   & 0.000 & 0.555 & 0.723   & 1 & \nodata &   & 2.245 & 8 \\
HD 104893    & 0.051 & 0.823 & 1.000   & 2 & 0.643   & 3 & 3.00 & 9 \\
HD 106373    & 0.044 & 0.338 & \nodata &   & \nodata &   & 1.51 & 8 \\
BD+4 2621    & 0.003 & 0.595 & 0.764   & 1 & \nodata &   & 2.43 & 9 \\
BD+1 2916    & 0.011 & 0.911 & 1.049   & 1 & 0.685   & 3 & 3.24 & 9 \\
HD 128279    & 0.039 & 0.469 & \nodata &   & 0.439   & 3 & 1.99 & 8 \\
HD 151559    & 0.069 & 0.584 & \nodata &   & \nodata &   & 2.40 & 9 \\
BD+17 3248   & 0.040 & 0.486 & 0.638   & 1 & \nodata &   & 2.04 & 8 \\
BD+25 3410   & 0.083 & 0.440 & \nodata &   & \nodata &   & 1.835 & 5 \\
HD 165195    & 0.099 & 0.920 & 1.076   & 1 & 0.710   & 4 & 3.30 & 8 \\
HD 166161    & 0.209 & 0.686 & 0.891   & 1 & 0.622   & 3 & 2.745 & 5 \\
HD 184711    & 0.081 & 0.938 & \nodata &   & 0.740   & 3 & 3.31 & 9 \\
HDE 232078   & 0.400 & 1.437 & 1.62    & 1 & \nodata &   & 4.50 & 9 \\
HD 187111    & 0.081 & 0.835 & 1.026   & 1 & 0.678   & 3 & 3.20 & 9 \\
HD 190287    & 0.026 & 0.503 & \nodata &   & 0.447   & 3 & 2.13 & 9 \\
HD 204543    & 0.024 & 0.635 & 0.817   & 1 & \nodata &   & 2.55 & 9 \\
HD 215601    & 0.009 & 0.532 & \nodata &   & 0.472   & 3 & 2.20 & 9 \\
HD 216143    & 0.013 & 0.690 & 0.874   & 1 & \nodata &   & 2.72 & 9 \\
HD 218620    & 0.007 & 0.720 & \nodata &   & \nodata &   & 2.76 & 9 \\
HD 218857    & 0.021 & 0.500 & 0.687   & 1 & \nodata &   & 2.12 & 9 \\
\enddata
\\
\tablerefs{(1) Stone 1983; (2) Carney 1980; (3) Norris, Bessell,
\& Pickles 1985; (4) Carney 1983; (5) Carney \& Latham 1986;
(6) Arribas \& Martinez Roger 1987; (7) Alonso, Arribas, \&
Martinez Roger 1994; (8) Alonso, Arribas, \& Martinez Roger 1998;
(9) 2MASS.}

\end{deluxetable}

\clearpage

\begin{deluxetable}{lcrrrrrrr}
\tablewidth{0pc}
\tablenum{2}
\tabletypesize{\footnotesize}
\tablecaption{Stellar Parameters \label{tab2}}
\tablehead{
\colhead{Star} &
\colhead{$\alpha$ (J2000) $\delta$} &
\colhead{[Fe/H]} &
\colhead{$M_{\rm V}$} &
\colhead{$T_{\rm eff}$} &
\colhead{log~$g$} &
\colhead{$R$/\rsun} &
\colhead{d(pc)} &
\colhead{Synthetic Spectrum} \\
\colhead{} & 
\colhead{} & 
\colhead{} & 
\colhead{} & 
\colhead{} &
\colhead{} & 
\colhead{} & 
\colhead{} & 
\colhead{$T_{\rm eff}$/log~$g$/[Fe/H]/$V_{\rm rot}$}   }
\startdata
HD 2665      & 00 30 45.4 $+57$ 03 53  & $-1.99$ & 0.66    & 5000 & 2.34 & 10.1 &  240 & 5000/2.5/$-$2.0/0 \\
HD 2796      & 00 31 16.9 $-16$ 47 40  & $-2.45$ & $-0.81$ & 4830 & 1.67 & 21.8 &  720 & 4750/1.5/$-$2.5/8 \\
HD 4306      & 00 45 27.1 $-09$ 32 39  & $-2.49$ & 0.19    & 4940 & 2.12 & 13.0 &  560 & 5000/2.0/$-$2.5/4 \\
HD 6268      & 01 03 18.1 $-27$ 52 50  & $-2.25$ & $-1.06$ & 4710 & 1.51 & 26.2 &  670 & 4750/1.5/$-$2.0/4 \\
HD 6229      & 01 03 36.4 $+23$ 46 06  & $-1.09$ & 0.84    & 5160 & 2.49 &  8.4 &  340 & 5250/2.5/$-$1.0/8 \\
HD 6833      & 01 09 52.2 $+54$ 44 20  & $-1.04$ & $-0.40$ & 4450 & 1.63 & 22.8 &  230 & 4500/1.5/$-$1.0/8 \\
HD 8724      & 01 26 17.5 $+17$ 07 35  & $-2.01$ & $-1.11$ & 4480 & 1.36 & 31.1 &  730 & 4500/1.5/$-$2.0/6 \\
HD 11582     & 01 53 00.2 $-34$ 17 36  & $-1.57$ & 1.88    & 5150 & 2.90 &  5.3 &  340 & 5000/3.0/$-$1.5/1 \\
HD 13979     & 02 15 20.8 $-25$ 54 54  & $-2.55$ & $-0.33$ & 4980 & 1.93 & 16.1 &  800 & 5000/2.0/$-$2.5/2 \\
CD$-36$ 1052 & 02 47 37.4 $-36$ 06 27  & $-2.19$ & 0.62    & 5980 & 2.69 &  6.7 &  720 & 6000/2.5/$-$2.0/16 \\
HD 21022     & 03 22 21.6 $-32$ 59 40  & $-1.17$ & $-1.17$ & 4500 & 1.35 & 31.4 & 1190 & 4500/1.5/$-$1.0/4 \\
CD$-24$ 1782 & 03 38 41.4 $-24$ 02 50  & $-2.66$ & 0.61    & 5150 & 2.38 &  9.6 &  730 & 5250/2.5/$-$2.5/1 \\
HD 25532     & 04 04 11.0 $+23$ 24 27  & $-1.33$ & 0.79    & 5320 & 2.54 &  8.0 &  270 & 5250/2.5/$-$1.5/8 \\
HD 26297     & 04 09 03.4 $-15$ 53 27  & $-1.67$ & $-1.48$ & 4280 & 1.08 & 42.9 &  620 & 4250/1.0/$-$1.5/6 \\
BD+6 648     & 04 13 13.1 $+06$ 36 01  & $-1.82$ & $-1.79$ & 4160 & 0.87 & 54.6 & 1250 & 4250/1.0/$-$2.0/6 \\
HD 29574     & 04 38 55.7 $-13$ 20 48  & $-1.63$ & $-2.11$ & 3960 & 0.57 & 77.3 & 1160 & 4000/0.5/$-$1.5/10 \\
HD 33771     & 05 10 49.6 $-37$ 49 03  & $-1.93$ & $-0.36$ & 4680 & 1.77 & 19.3 &  890 & 4750/1.5/$-$2.0/6 \\
HD 41667     & 06 05 03.6 $-32$ 59 39  & $-1.18$ & +0.07 & 4580 & 1.89 & 16.8 &  490 & 4500/2.0/$-2.0$/4 \\
HD 44007     & 06 18 48.4 $-14$ 50 43  & $-1.23$ & 1.83    & 4980 & 2.80 &  5.9 &  150 & 5000/3.0/$-$1.0/0 \\
HD 74462     & 08 48 20.6 $+67$ 26 59  & $-1.60$ & $-0.84$ & 4510 & 1.49 & 26.8 &  790 & 4500/1.5/$-$1.5/4 \\
BD+54 1323   & 09 42 19.4 $+53$ 28 26  & $-1.85$ & 0.69    & 5120 & 2.41 &  9.3 &  530 & 5000/2.5/$-$2.0/4 \\
BD+58 1218   & 09 52 38.6 $+57$ 54 58  & $-2.84$ & 0.39    & 4950 & 2.20 & 11.8 &  820 & 5000/2.0/$-$3.0/4 \\
HD 85773     & 09 53 39.2 $-22$ 50 08  & $-2.22$ & $-1.97$ & 4300 & 0.90 & 53.0 & 1820 & 4250/1.0/$-$2.0/6 \\
HD 88609     & 10 14 28.9 $+53$ 33 39  & $-2.58$ & $-1.42$ & 4470 & 1.22 & 36.5 & 1010 & 4500/1.0/$-$2.5/4 \\
HD 101063    & 11 37 40.4 $-28$ 51 04  & $-1.09$ & 2.74    & 5070 & 3.21 &  3.7 &  210 & 5000/3.0/$-$1.0/2 \\
BD+52 1601   & 11 59 59.0 $+51$ 46 17  & $-1.49$ & 0.13    & 4800 & 2.03 & 14.3 &  540 & 4750/2.0/$-$1.5/4 \\
HD 104893    & 12 04 43.1 $-29$ 11 05  & $-1.78$ & $-1.70$ & 4270 & 0.99 & 47.6 & 1400 & 4250/1.0/$-$2.0/2 \\
HD 106373    & 12 14 13.3 $-28$ 15 06  & $-2.48$ & 0.57    & 6160 & 2.71 &  6.5 &  430 & 6250/2.5/$-$2.5/12 \\
BD+4 2621    & 12 28 44.6 $+04$ 01 26  & $-2.51$ & $-0.87$ & 4710 & 1.58 & 24.1 & 1460 & 4750/1.5/$-$2.5/2 \\
BD+1 2916    & 14 21 45.1 $+00$ 46 59  & $-1.45$ & $-1.76$ & 4040 & 0.78 & 60.1 & 1920 & 4000/0.5/$-$1.5/8 \\
HD 128279    & 14 36 48.5 $-29$ 06 46  & $-1.97$ & 1.86    & 5290 & 2.94 &  5.0 &  160 & 5250/3.0/$-$2.0/0 \\
HD 151559    & 16 49 08.1 $-27$ 34 15  & $-1.01$ & 1.65    & 4970 & 2.73 &  6.4 &  250 & 5000/2.5/$-$1.0/0 \\
BD+17 3248   & 17 28 14.4 $+17$ 30 35  & $-2.07$ & 0.65    & 5240 & 2.44 &  9.0 &  510 & 5250/2.5/$-$2.0/4 \\
BD+25 3410   & 18 02 03.2 $+25$ 00 41  & $-1.37$ & 0.79    & 5740 & 2.69 &  6.7 &  450 & 5750/2.5/$-$1.5/10 \\
HD 165195    & 18 04 40.0 $+03$ 46 44  & $-2.16$ & $-2.14$ & 4200 & 0.76 & 61.9 &  640 & 4250/0.5/$-$2.0/8 \\
HD 166161    & 18 09 40.6 $-08$ 46 45  & $-1.33$ & 0.79    & 5070 & 2.43 &  9.0 &  190 & 5000/2.5/$-$1.5/4 \\
HD 184711    & 19 37 11.9 $-39$ 44 37  & $-2.29$ & $-2.35$ & 4090 & 0.59 & 75.4 & 1000 & 4000/0.5/$-$2.5/6 \\
HDE 232078   & 19 38 12.1 $+16$ 48 26  & $-1.54$ & $-2.15$ & 4000 & 0.57 & 76.7 &  660 & 4000/0.5/$-1.5$/10 \\
HD 187111    & 19 48 39.5 $-12$ 07 19  & $-1.65$ & $-1.54$ & 4260 & 1.04 & 44.6 &  610 & 4250/1.0/$-$1.5/4 \\
HD 190287    & 20 05 38.2 $-34$ 55 10  & $-1.09$ & 2.63    & 5090 & 3.18 &  3.8 &  150 & 5000/3.0/$-$1.0/2 \\
HD 204543    & 21 29 28.2 $-03$ 30 55  & $-1.85$ & $-1.09$ & 4610 & 1.44 & 28.1 &  720 & 4500/1.5/$-$2.0/6 \\
HD 215601    & 22 46 48.0 $-31$ 52 18  & $-1.56$ & $-0.17$ & 4865 & 1.95 & 15.8 &  530 & 4750/2.0/$-$1.5/4 \\
HD 216143    & 22 50 31.0 $-06$ 54 49  & $-2.01$ & $-1.42$ & 4445 & 1.21 & 36.8 &  690 & 4500/1.0/$-$2.0/6 \\
HD 218620    & 23 09 40.1 $-30$ 54 36  & $-1.55$ & $-1.22$ & 4370 & 1.25 & 35.3 & 1300 & 4250/1.0/$-$1.5/4 \\
HD 218857    & 23 11 24.6 $-16$ 15 04  & $-2.15$ & 0.81    & 5040 & 2.42 &  9.2 &  410 & 5000/2.5/$-$2.0/2 \\  
\enddata
\end{deluxetable}

\clearpage

\begin{deluxetable}{llrr}
\tablewidth{0pc}
\tablenum{3}
\tablecaption{Radial Velocities \label{tab3}}
\tablehead{
\colhead{Telescope} & \colhead{HJD} &
\colhead{$V_{\rm rad}$}  & \colhead{$\sigma_{\rm int}$}}
 
\startdata 
\cutinhead{HD 20~~~~00 05 15.3 -27 16 18}
M \dotfill & 2446006.67547 & $-57.75$ & 0.35 \\
T \dotfill & 2446361.77031 & $-56.78$ & 0.29 \\
T \dotfill & 2446369.79985 & $-56.58$ & 0.22 \\
M \dotfill & 2446393.62169 & $-58.04$ & 0.42 \\
M \dotfill & 2446398.57495 & $-57.44$ & 0.25 \\
M \dotfill & 2446725.71814 & $-57.31$ & 0.34 \\
T \dotfill & 2447044.86526 & $-56.77$ & 0.25 \\
M \dotfill & 2447345.98195 & $-57.01$ & 0.26 \\
M \dotfill & 2447696.97977 & $-57.25$ & 0.25 \\
T \dotfill & 2448141.87887 & $-57.03$ & 0.34 \\
T \dotfill & 2449195.95998 & $-57.25$ & 0.41 \\
T \dotfill & 2449265.79665 & $-56.87$ & 0.39 \\
M \dotfill & 2450647.99077 & $-57.34$ & 0.32 \\
\tablerefs{M = MMT; T = Tillinghast reflector, Mt.\ Hopkins
Observatory; W = Wyeth reflector, Oak Ridge Observatory}
\enddata
\end{deluxetable}

\clearpage

\begin{deluxetable}{lrrrrrrrrrr}

\tablewidth{0pc}
\tablenum{4}
\tablecaption{Radial Velocity Summary $-$ Prior Study \label{tab4}}
\footnotesize
\tablehead{
\colhead{Star} &
\colhead{N} &
\colhead{Span} &
\colhead{$<V_{\rm rad}>$} &
\colhead{$\sigma$($V_{\rm rad}$)} &
\colhead{E} &
\colhead{I} &
\colhead{E/I} &
\colhead{$\chi^{2}$} &
\colhead{P($\chi^{2}$)} &
\colhead{$V_{\rm broad}$}}
\startdata
HD 97  &    42 &   7687 &    75.64 &   0.13 &   0.87 &   0.59 &   1.47 &   78.61 & 0.000370 &    3.8  \\
HD 3179  &    20 &   7503 &   $-$74.73 &   0.22 &   0.97 &   0.64 &   1.51 & 37.17 & 0.007559 &    5.5  \\
HD 6755  &    36 &   7713 &  $-$318.36 &   0.12 &   5.96 &   0.42 &  14.17 & 6676.71 & 0.000000 &    3.3  \\
HD 27928  &    38 &   6830 &    44.40 &   0.17 &   2.51 &   0.43 &   5.83 & 1188.63 & 0.000000 &    0.0  \\
BD$-$1 2582 &    52 &   7938 &    0.35 &   0.19 &   1.34 &   0.68 &   1.97 &  232.78 & 0.000000 &    6.6  \\
HD 108317  &    48 &   4573 &     7.27 &   0.15 &   1.03 &   0.56 &   1.82 & 165.77 & 0.000000 &    5.1  \\
BD+3 2782 &    14 &   7363 &    31.97 &   0.16 &   0.60 &   0.52 &   1.16 & 17.11 & 0.194256 &    5.0  \\
HD 121261  &    15 &   7684 &   100.03 &   0.22 &   0.85 &   0.44 &   1.92 & 61.15 & 0.000000 &    7.7  \\
BD+18 2976 &   106 &   7528 &  $-$167.97 &   0.09 &   3.54 &   0.55 &   6.47 & 4915.69 & 0.000000 &    5.1  \\
BD+1 3070 &    43 &   7731 &  $-$330.11 &   0.11 &   2.86 &   0.66 &   4.34 &   918.75 & 000000 &    4.9  \\
HD 213467  &    14 &   7627 &   $-$72.50 &   0.19 &   0.71 &   0.44 &   1.59 & 42.87 & 0.000047 &    1.8  \\
\\
\enddata
\end{deluxetable}

\clearpage

\begin{deluxetable}{lrrrrrrrrrr}

\tablewidth{0pc}
\tablenum{5}
\tablecaption{Radial Velocity Summary $-$ New Program \label{tab5}}
\footnotesize
\tablehead{
\colhead{Star} &
\colhead{N} &
\colhead{Span} &
\colhead{$<V_{\rm rad}>$} &
\colhead{$\sigma$($V_{\rm rad}$)} &
\colhead{E} &
\colhead{I} &
\colhead{E/I} &
\colhead{$\chi^{2}$} &
\colhead{P($\chi^{2}$)} &
\colhead{$V_{\rm broad}$}}
\startdata
HD 2665   &     7 &   8066 &  $-382.68$ &   0.22 &   0.59 &   0.47 &   1.26 &    15.20 & 0.018728 &    0.5  \\
HD 2796   &     5 &    763 &  $ -60.99$ &   0.24 &   0.51 &   0.53 &   0.96 &     2.45 & 0.654283 &    7.0  \\
HD 4306   &    14 &   1242 &  $ -66.48$ &   1.38 &   5.17 &   0.53 &   9.75 &    1334.63 & 0.000000 &    0.9  \\
HD 6268   &     6 &    745 &    38.68 &   0.24 &   0.58 &   0.40 &   1.40 &     10.65 & 0.058702 &    4.3  \\
HD 6229   &     6 &    736 &  $ -90.88$ &   0.17 &   0.36 &   0.41 &   0.87 &     3.72 & 0.589905 &    7.0  \\
HD 6833   &    34 &   7755 & $ -243.14$ &   0.11 &   0.62 &   0.33 &   1.89 &   116.20 & 0.000000 &    7.4  \\
HD 8724   &     7 &    811 & $ -113.19$ &   0.20 &   0.53 &   0.45 &   1.18 &     9.59 & 0.143010 &    6.0  \\
HD 11582  &     6 &    745 &    21.53 &   0.19 &   0.48 &   0.41 &   1.16 &     7.42 & 0.191459 &    0.5  \\
HD 13979  &     5 &    453 &    51.62 &   0.21 &   0.15 &   0.48 &   0.31 &     0.39 & 0.983112 &    3.7  \\
CD$-36$ 1052 &     5 &    741 &   304.59 &   0.30 &   0.36 &   0.68 &   0.53 &     1.11 & 0.893207 &   14.2  \\
HD 21022  &     5 &    749 &   122.29 &   0.26 &   0.59 &   0.55 &   1.07 &     4.90 & 0.297427 &    5.0  \\
CD$-24$ 1782 &     6 &    741 &   101.43 &   0.26 &   0.63 &   0.50 &   1.25 &     9.00 & 0.109119 &    0.5  \\
HD 25532  &     5 &    560 &  $-112.24$ &   0.20 &   0.39 &   0.44 &   0.89 &     3.27 & 0.513139 &    8.3  \\
HD 26297  &     5 &    419 &    14.83 &   0.25 &   0.56 &   0.43 &   1.31 &     5.59 & 0.231752 &    5.6  \\
BD+6 648 &     4 &    403 &  $-143.46$ &   0.20 &   0.27 &   0.41 &   0.65 &     1.30 & 0.728639 &    6.1  \\
HD 29574  &    22 &   7427 &    19.80 &   0.31 &   1.46 &   0.59 &   2.45 & 
132.26 & 0.000000 &   10.1  \\
HD 33771  &     4 &    594 &   $-14.24$ &   0.19 &   0.37 &   0.39 &   0.97 &     3.04 & 0.385667 &    5.0  \\
HD 41667  &     3 &    356 &  296.73  &  0.21 & 0.36 & 0.36 & 1.01 & 1.94 &     0.379803 & 5.0 \\
HD 44007  &     5 &    387 &   161.83 &   0.23 &   0.33 &   0.52 &   0.62 &     1.63 & 0.804223 &    0.5  \\
HD 74462  &     7 &    745 &  $-169.03$ &   0.15 &   0.39 &   0.36 &   1.10 &     7.51 & 0.276206 &    4.6  \\
BD+54 1323 &     5 &    771 &   $-67.11$ &   0.25 &   0.52 &   0.56 &   0.94 &     4.42 & 0.352620 &    3.0  \\
BD+58 1218 &     5 &   1013 &  $-304.12$ &   0.31 &   0.36 &   0.70 &   0.51 &     1.09 & 0.895896 &    1.5  \\
HD 85773  &     5 &    794 &   147.41 &   0.23 &   0.41 &   0.52 &   0.79 &     1.09 & 0.599917 &    4.2  \\
HD 88609  &     5 &    738 &   $-38.49$ &   0.23 &   0.40 &   0.51 &   0.78 &     2.27 & 0.686534 &    4.0  \\
HD 101063  &    16 &   1036 &   182.34 &   0.21 &   0.86 &   0.52 &   1.64 &    26.72 & 0.031118 &    3.8  \\
BD+52 1601 &     6 &   1037 &   $-46.95$ &   0.17 &   0.29 &   0.42 &   0.67 &     2.45 & 0.784585 &    4.6  \\
HD 104893  &     8 &    623 &    24.12 &   0.16 &   0.24 &   0.44 &   0.55 &     2.12 & 0.953138 &    3.0  \\
HD 106373  &     7 &   1062 &    83.67 &   0.23 &   0.40 &   0.61 &   0.66 &     2.89 & 0.822853 &   13.7  \\
BD+4 2621 &     6 &   1011 &   $-41.96$ &   0.26 &   0.30 &   0.63 &   0.47 &     1.32 & 0.932883 &    2.5  \\
BD+1 2916 &    14 &   1003 &   $-12.10$ &   0.22 &   0.82 &   0.59 &   1.39 &    25.22 & 0.021587 &    8.2  \\
HD 128279  &     7 &    616 &   $-76.54$ &   0.23 &   0.61 &   0.44 &   1.38 &     9.85 & 0.131045 &    0.5  \\
HD 151559  &     4 &    377 &    16.48 &   0.19 &   0.35 &   0.39 &   0.91 &     2.41 & 0.492020 &    0.0  \\
BD+17 3248 &     6 &   1091 &  $-146.87$ &   0.23 &   0.37 &   0.55 &   0.67 &     2.30 & 0.806028 &    3.8  \\
BD+25 3410 &     6 &   1095 &  $-187.76$ &   0.32 &   0.67 &   0.77 &   0.87 &     4.77 & 0.444779 &    9.7  \\
HD 165195   &    16 &   7714 &    $-0.33$ &   0.25 &   1.00 &   0.46 &   2.15 &    69.53 & 0.000000 &    7.6  \\
HD 166161  &     6 &    900 &    68.33 &   0.18 &   0.32 &   0.43 &   0.74 &     3.21 & 0.667236 &    4.0  \\
HD 184711  &     6 &    475 &   102.21 &   1.29 &   3.15 &   0.49 &   6.39 &   1236.97 & 0.000000 &    7.7  \\
HDE 232078 &    19 &   7917 &  $-387.16$ &  0.34 &   1.50 &   0.54 &   2.77  &  168.72 & 0.000000 &   10.9 \\
HD 187111  &     5 &    781 &  $-186.48$ &   0.18 &   0.11 &   0.39 &   0.28 &     0.31 & 0.989185 &    5.2  \\
HD 190287  &     5 &    538 &   143.32 &   0.19 &   0.38 &   0.44 &   0.87 &     3.22 & 0.521124 &    3.4  \\
HD 204543  &     7 &   7570 &   $-98.35$ &   0.17 &   0.37 &   0.45 &   0.83 &     6.86 & 0.334038 &    5.0  \\
HD 215601  &     5 &    749 &   $-37.14$ &   0.21 &   0.48 &   0.38 &   1.25 &     6.73 & 0.150808 &    5.0  \\
HD 216143  &     5 &    742 &  $-115.99$ &   0.23 &   0.51 &   0.43 &   1.18 &     5.72 & 0.221185 &    5.5  \\
HD 218620  &     5 &    749 &   $-52.34$ &   0.33 &   0.73 &   0.38 &   1.91 &    16.62 & 0.002294 &    5.2  \\
HD 218857  &     9 &   7648 &  $-170.39$ &   0.27 &   0.82 &   0.54 &   1.54 &    18.91 & 0.015356 &    2.1  \\
\\
\enddata
\end{deluxetable}

\clearpage

\clearpage

\clearpage

\clearpage

\clearpage

\clearpage

\clearpage

\clearpage

\clearpage

\clearpage

\clearpage

\clearpage

\clearpage

\begin{references}

\reference{aamr94} Alonso, A.,
Arribas, S., \& Martinez Roger, C.\ 1994, \aaps, 107, 365

\reference{aamr98} Alonso, A.,
Arribas, S., \& Martinez Roger, C.\ 1998, \aaps, 131, 209

\reference{aamr99} Alonso, A.,
Arribas, S., \& Martinez Roger, C.\ 1999, \aaps, 140, 261

\reference{aamr01} Alonso, A.,
Arribas, S., \& Martinez Roger, C.\ 2001, \aap, 376, 1039

\reference{att94} 
Anthony-Twarog, B.\ J., \& Twarog, B.\ A.\ 1994, \aj, 107, 1577 (ATT)

\reference{amr}
Arribas, S., \& Martinez Roger, C.\ 1987,
\aaps, 70, 303

\reference{behr2003} Behr, B.\ 2003, \apjs, 149, 101

\reference{behr2000a} Behr, B.\ B., Cohen, J.\ G., \& McCarthy, J.\ K.\ 2000,
\apj, 531, L37

\reference{behr2000b} Behr, B.\ B., Djorgovski, S.\ G., Cohen, J.\ G.,
McCarthy, J.\ K., C\^{o}t\'{e}, P., Piotto, G., \& Zoccalo, M.\ 2000,
\apj, 528, 849

\reference{bond80} Bond, H.\ E.\ 1980, 
\apjs, 44, 517

\reference{bopp81} Bopp, B.\ W., \& Stencel, R.\ E.\ 1981, \apj, 247, L31

\reference{bw} Brown, J.\ A.,
\& Wallerstein, G.\ 1992, \aj, 104, 1818

\reference{burris2000} Burris, D.\ L., Pilachowski, C.\ A., Armandroff,
T.\ E., Sneden, C., Cowan J.\ J., \& Roe, H.\ 2000, \apj, 544, 302
 
\reference{c80} Carney, B.\ W.\ 1980,
\aj, 85, 38

\reference{c83} Carney, B.\ W.\ 1983,
\aj, 88, 623

\reference{carney2007} Carney, B.\ W., Gray, D.\ F.,
Yong, D., Latham, D.\ W., Manset, N., Zelman, R., \&
Laird, J.\ B.\ 2007, in preparation

\reference{cl86} Carney, B.\ W.,
\& Latham, D.\ W.\ 1986, \aj, 92, 60 (C\&L)

\reference{c2003} Carney, B.\ W., Latham, D.\ W., Stefanik, R.\ P.,
Laird, J.\ B., \& Morse, J.\ A.\ 2003, \aj, 125, 293 (C2003)

\reference{cohen2000} Cohen, J.\ G., \& McCarthy, J.\ K.\ 1997,
\aj, 113, 1353

\reference{cote02} C\^{o}t\'{e}, Djorgovski, S.\ G., Meylan, G.,
Castro, S., \& McCarthy, J.\ K.\ 2002, \apj, 574, 782

\reference{cote96} C\^{o}t\'{e}, P., Pryor, C., McClure, R.\ D.,
Fletcher, J.\ M., \& Hesser, J.\ E.\ 1996, \aj, 112,574

\reference{demedeiros00} de Medeiros, J.\ R., do~Nascimento, Jr.,
J.\ D., Sankarankutty, S., Costa, J.\ M., \& Maia, M.\ R.\ G.\ 2000,
\aap, 363, 239

\reference{demedeiros06} de Medeiros, J.\ R., Silva, J.\ R.\ P.,
Do~Nascimento, J.\ D., Jr., Canto~Martins, B.\ L., da~Silva, L.,
Melo, C., \& Burnet, M.\ 2006, \aap, 458, 895

\reference{fernie83} Fernie, J.\ D.\ 1983, \pasp, 95, 782

\reference{fulbright} Fulbright, J.\ A.\ 2000, \aj, 120, 1841

\reference{paper16} Goldberg, D.,
Mazeh, T., Latham, D.\ W., Stefanik, R.\ P., Carney, B.\ W., 
\& Laird, J.\ B.\ 2002, 
\aj, 124, 1132

\reference{gratton1996} Gratton, R.\ G., Carretta, E., \& Castelli, F.\ 1996,
\aap, 314, 191

\reference{gray82} Gray, D.\ F.\ 1982, \apj, 262, 282

\reference{gray84} Gray, D.\ F.\ 1984, \apj, 281, 719

\reference{gray89} Gray, D.\ F., \& Pallavicini, R.\ 1989, \pasp, 101, 685

\reference{gray86} Gray, D.\ F., \& Toner, C.\ G.\ 1986, \apj, 310, 277

\reference{gray87} Gray, D.\ F., \& Toner, C.\ G.\ 1987, \apj, 322, 360

\reference{gunngriffin} Gunn, J.\ E., \& Griffin, R.\ F.\ 1979, \aj, 84, 752

\reference{kinman2000} Kinman, T., Castelli, F., Cacciari, C.,
Bragaglia, A., Harmer, D., \& Valdes, F.\ 2000, \aap, 364, 102

\reference{kraftivans} Kraft, R.\ P., \& Ivans, I.\ I.\ 2003, \pasp,
115, 143

\reference{kraft92} Kraft, R.\ P.,
Sneden, C., Langer, G.\ E., \& Prosser, C.\ F.\ 1992, \aj,
104, 645

\reference{kraft97} Kraft, R.\ P.,
Sneden, C., Smith, G.\ H., Shetrone, M.\ D., Langer, G.\ E.,
\& Pilachowski, C.\ A.\ 1997, \aj, 113, 279

\reference{kurtz} Kurtz, M.\ J., \& Mink, D.\ J.\ 1998,
\pasp, 110, 934

\reference{kurucz84} Kurucz, R.\ L., Furenlid, I. Brault, J.,
\& Testerman, L.\ 1984. {\em Solar Flux Atlas from 296 to 1300 nm},
(National Solar Observatory, Sunspot, NM)

\reference{latham85} Latham, D.\ W.\ 1985,
in Stellar Radial Velocities, IAU Coll.\ No.\ 88, ed.\ A.\ G.\ D.\
Philip \& D.\ W.\ Latham (L.\ Davis Press, Schenectady), p.\ 5

\reference{latham92} Latham, D.\ W.\ 1992,
in Complementary Approaches to Binary and Multiple Star
Research, IAU Coll.\ No.\ 135, ed.\ H.\ McAlister \& W.\
Hartkopf (ASP, San Francisco), p.\ 110

\reference{paper14} Latham, D.\ W.,
Stefanik, R.\ P., Torres, G., Davis, R.\ J., Mazeh, T., Carney, B.\ W.,
Laird, J.\ B., \& Morse, J.\ A.\ 2002,
\aj, 124, 1144

\reference{lupton87} Lupton, R.\ H., Gunn, J.\ E., \& Griffin, R.\ F.\ 1987,
\aj, 93, 114

\reference{massarotti07} Massarotti, A., Latham D.\ W., Stefanik,
R.\ P., Fogel, J., Carney, B.\ W., \& Laird, J.\ B.\ 2007, in preparation

\reference{mayormermilliod84} Mayor, M.,
\& Mermilliod, J.-C.\ 1984, in Observational Tests of
Stellar Evolution Theory, IAU Sym.\ No.\ 105, ed.\ A.\ Maeder \&
A.\ Renzini (Reidel, Dordrecht), p.\ 411
 
\reference{mayor97} Mayor, M.,
Meylan, G., Udry, S., Duquennoy, A., Andersen, J.,
Nordstr\"{o}m, B., Imbert, M., Maurice, M.,
Pr\'{e}vot, L., Ardeberg, A., \& Lindgren, H.\ 1997,
\aj, 114, 1087

\reference{nordstrom} Nordstr\"{o}m,
B., Latham, D.\ W., Morse, J.\ A., Milone, A.\ A.\ E., Kurucz, R.\ L.,
Andersen, J., \& Stefanik, R.\ P.\ 1994, \aap, 287, 338

\reference{nbp} Norris, J.,
Bessell, M.\ S., \& Pickles, A.\ J.\ 1985, \apjs, 58, 463 (NBP)

\reference{peterson1983} Peterson, R.\ C.\ 1983, \apj, 275, 737
 
\reference{peterson1985a} Peterson, R.\ C.\ 1985, \apj, 289, 320
 
\reference{peterson95} Peterson, R.\ C., Rood, R.\ T.,
\& Crocker, D.\ A.\ 1995, \apj, 453, 214

\reference{peterson83} Peterson, R.\ C., Tarbell, T.\ D., \&
Carney, B.\ W.\ 1983, \apj, 265, 972

\reference{pryor88} Pryor, C., Latham D.\ W., \& Hazen, M.\ L.\ 1988,
\aj, 96, 123

\reference{recioblanco} Recio-Blanco, A., Piotto, G., Aparicio, A.,
\& Renzini, A.\ 2002, astro-ph 0204403

\reference{sl99} Siess, L., \& Livio, M.\ 1999, \mnras, 308, 1133

\reference{skrutskie2006} Skrutskie, M.\ F.\ et al.\ 2006, \aj, 131, 1163

\reference{soker98} Soker, N.\ 1998, \aj, 116, 1308

\reference{sozzetti06} Sozzetti, A., Torres, G., Latham, D.\ W.,
Carney, B.\ W., Stefanik, R.\ P., Boss, A.\ P., \& Laird, J.\ B.\
2006, \apj, 649, 428

\reference{stefanik1999} Stefanik, R.\ P., Latham, D.\ W., \&
Torres, G.\ 1999, in Precise Radial Velocities, ASP Conf. Series No. 185,
ed. J.\ B.\ Hearnshaw \& C.\ D.\ Scarfe (ASP, San Francisco), p.\ 354

\reference{stone83} Stone, R.\ P.\ S.\ 1983,
\pasp, 95, 27

\reference{straniero91} Straniero,
O., \& Chieffi, A.\ 1991, \apjs, 76, 525

\end{references}
\end{document}